\documentclass[11pt,a4paper]{article}
\usepackage{comment}
\usepackage{graphicx}
\usepackage{amssymb}
\usepackage{amsmath}
\usepackage{amsfonts}
\usepackage{dsfont}
\usepackage{mathtools}
\usepackage{array}
\usepackage{soul}
\usepackage{float}
\usepackage{rotating}
\usepackage{bbold,amsfonts}
\allowdisplaybreaks

\usepackage[utf8]{inputenc}
\usepackage{bm}
\usepackage{xcolor}
\usepackage{float}
\usepackage{braket}
\usepackage{placeins}
\usepackage[height=8.8in,width=6.45in]{geometry}
\usepackage[font=small,labelfont=bf]{caption}
\usepackage[hidelinks]{hyperref}
\usepackage{booktabs,float,slashed}
\usepackage{standalone}
\usepackage{caption}
\usepackage{subcaption}
\usepackage{makecell}
\usepackage[maxbibnames=99,sorting=none,giveninits=true,backend=biber,style=numeric-comp,sortcites,doi=false,hyperref=true]{biblatex}
\renewbibmacro{in:}{}
\usepackage{hyperref}
\DeclareFieldFormat{doilink}{\iffieldundef{doi}{#1}{\href{https://doi.org/\thefield{doi}}{#1}}}
\DeclareFieldFormat[article,periodical]{volume}{\mkbibbold{#1}}
\DeclareFieldFormat[article,periodical]{journaltitle}{#1}
\DeclareFieldFormat[article,periodical]{pages}{#1}

\usepackage{xpatch}
\xpatchbibdriver{article}
  {\usebibmacro{journal+issuetitle}%
   \newunit
   \usebibmacro{byeditor+others}%
   \newunit
   \usebibmacro{note+pages}}
  {\printtext[doilink]{%
     \usebibmacro{journal+issuetitle}%
     \newunit
     \usebibmacro{byeditor+others}%
     \newunit
     \usebibmacro{note+pages}}}
  {}{}
\numberwithin{equation}{section}
\newcommand{\beq}{\begin{equation}}
\newcommand{\eeq}{\end{equation}}

\newcommand {\non}{\nonumber}

\newcommand{\p}{\pi}

\makeatletter
\newcommand*{\letterdef@}{}
\newcommand*{\letterdef}[3]{%
	\def\letterdef@##1{\expandafter\newcommand\csname #1\endcsname{#2{##1}}}%
	\@tfor\@tempa :=#3\do{\expandafter\letterdef@\expandafter{\@tempa}}}
\makeatother
\letterdef{c#1} {\mathcal}{ABCDEFGHIJKLMNOPQRSTUVWXYZ} % \cX = \mathcal{X}
\letterdef{rm#1}{\mathrm} {dDeimM} % \rmX = \mathrm{X} for X in {dDmM}

\title{Draft-Wilson lines in $Z_2$ quiver theory}

\begin{document}

%\maketitle
\begin{titlepage}

\begin{flushright}
\small
\texttt{HU-EP-26/08}
\end{flushright}

\vspace*{10mm}
\begin{center}
{ \LARGE \bf Non-perturbative corrections to line defect\\ integrated correlators in $Sp(N)$ SCFTs
}

\vspace*{15mm}

{\Large Lorenzo De Lillo${}^{\,a,b}$ and  Alessandro Pini${}^{\,c}$}

\vspace*{8mm}

${}^a$ Universit\`a di Torino, Dipartimento di Fisica,\\
			Via P. Giuria 1, I-10125 Torino, Italy
			\vskip 0.3cm
			
${}^b$   I.N.F.N. - sezione di Torino,\\
			Via P. Giuria 1, I-10125 Torino, Italy 
			\vskip 0.3cm

${}^c$ Institut f{\"u}r Physik, Humboldt-Universit{\"a}t zu Berlin,\\
     IRIS Geb{\"a}ude, Zum Großen Windkanal 2, 12489 Berlin, Germany  
     \vskip 0.3cm

\vskip 0.8cm
	{\small
		E-mail:
		\texttt{lorenzo.delillo@unito.it;alessandro.pini@physik.hu-berlin.de}
	}
\vspace*{0.8cm}
\end{center}

\begin{abstract}
We consider the $\mathcal{N}=4$ SYM theory with gauge group $Sp(N)$ and the $\mathcal{N}=2$ superconformal field theory consisting of four hypermultiplets in the fundamental representation and one hypermultiplet in the rank-two antisymmetric representation of the $Sp(N)$ gauge group. Building on previous results obtained via supersymmetric localization and a Toda equation, we determine the leading non-perturbative  corrections at strong coupling to the integrated correlator between a Wilson line and two Higgs-branch moment map operators. In the case of the $\mathcal{N}=2$ SCFT, the presence of truncated asymptotic expansions led us to develop a resurgent method complementary to Cheshire cat resurgence. This approach has the advantage of yielding an exact expression for the correlator in terms of an analytic function, which can subsequently be expanded in the strong-coupling regime.
\end{abstract}

\end{titlepage}
\setcounter{tocdepth}{2}

\newpage

\tableofcontents

\vspace*{1cm}

\section{Introduction}
In recent years, the study of four-dimensional Superconformal Field Theories (SCFTs) has attracted increasing attention, leading to many exact results obtained both in the $\mathcal{N}=4$ SYM maximally supersymmetric theories and in $\mathcal{N}=2$ SCFTs. This progress has been made possible by the use of powerful techniques such as integrability, the AdS/CFT correspondence, and supersymmetric localization \cite{Pestun:2007rz}. The latter, upon placing the gauge theory on the four-sphere $\mathbb{S}^4$, allows one to recast the evaluation of the infinite-dimensional path integral into the computation of an interacting matrix model. In this way many exact expressions have been obtained for a large class of observables both at finite $N$ and in the large $N$ limit of the gauge theory. Unlike the standard perturbative approach based on Feynman diagram computations, in many cases supersymmetric localization allows one to compute the perturbative series of a given observable to very high orders in the loop expansion. This, in turn, has made it possible to show that, the resulting perturbative series may have a finite radius of convergence and in some cases can be resummed. A well-known example is provided by the v.e.v. of the one-half BPS Wilson loop in $\mathcal{N}=4$ SYM \cite{Erickson:2000af}. Nevertheless, the situation is generally very different in the strong-coupling regime, i.e. when the gauge coupling constant $g$ is very large. In this case, the perturbative strong-coupling expansion of a given observable is typically an asymptotic and non-Borel-summable series, due to the presence of Borel singularities on the positive real axis of the Borel plane. These singularities are problematic because, when computing the inverse Borel transform, one must deform the integration contour to avoid them. This procedure introduces an ambiguity of order $\text{e}^{-A\,g}$,
where $A$ denotes the location of the singularity closest to the origin, reflecting the fact that different choices of integration contour lead to inequivalent results. Remarkably, this problem admits a beautiful resolution: the observable receives non-perturbative corrections of precisely the same order as the ambiguity mentioned above. These contributions cancel the contour ambiguity, rendering the full observable well defined and effectively Borel summable. In this way, the mathematical requirement of having well-defined, i.e. Borel summable, observables can lead to the discovery of genuinely non-perturbative physical effects. This mechanism, in which the ambiguities of the perturbative series are canceled by non-perturbative contributions, is known as \textit{Resurgence}. The mathematical theory was originally developed in \cite{Ecalle1981}; for reviews and more introductory expositions, see also \cite{Dorigoni:2014hea,marino_resurgence_course,Aniceto:2018bis,Dunne:2025mye}.
Moreover, for SCFTs with a known gravity dual, the interest in these results is not limited to the quantum field theory side, as they also admit an interpretation in the corresponding gravity dual in terms of \textit{worldsheet instantons}, providing potential predictions for future string theory computations, see for example \cite{Drukker:2006ga}. Finally, it is also worth mentioning that there are also observables, for instance the correlators considered in \cite{Brown:2025huy}, whose perturbative strong-coupling expansions contain only a finite number of terms. Importantly, even in these cases a technique has been developed, known as ``\textit{Cheshire cat resurgence}'', which allows one to uncover the corresponding non-perturbative corrections (see for example \cite{Dorigoni:2017smz,Kozcaz:2016wvy}).

The asymptotic expansions of many observables have been analyzed using resurgence techniques. These studies include, for example, the cusp anomalous dimension in $\mathcal{N}=4$ SYM \cite{Dorigoni:2015dha,Basso:2007wd,Beisert:2006ez,Dunne:2025wbq,Aniceto:2015rua}, v.e.v of half-BPS Wilson loops and 2-point function among chiral operators \cite{Bajnok:2024ymr,Bajnok:2024bqr,Bajnok:2025lji,Bajnok:2024epf}. Finally, for the scope of this article it is also worth mentioning a particularly interesting class of observables whose resurgence properties have been extensively analyzed: the integrated correlators of local scalar operators belonging to the stress-tensor multiplet in $\mathcal{N}=4$ \cite{Binder:2019jwn,Chester:2019jas,Dorigoni:2021bvj,Alday:2021vfb,Dorigoni:2024dhy,Chester:2025kvw,Chester:2019pvm}. These observables originate from four-point functions of local scalar operators belonging to the stress-tensor multiplet, which are then further integrated using a measure that is entirely fixed by superconformal symmetry, thereby eliminating any residual dependence on the space–time coordinates. Remarkably, in this integrated form, they can be computed exactly via supersymmetric localization. Although integrated correlators contain less information than their unintegrated counterparts, they nevertheless retain a substantial amount of physical information. More recently, these observables have also been studied within the context of some $\mathcal{N}=2$ SCFTs \cite{Chester:2022sqb,Fiol:2023cml,Behan:2023fqq,Billo:2023kak,Pini:2024uia,Billo:2024ftq,DeLillo:2025stg}.

In this work, we consider another type of integrated correlator, namely one involving the insertion of a defect operator \cite{Pufu:2023vwo}, whose resurgence properties, to the best of our knowledge, have not been investigated so far. In $\mathcal{N}=4$ SYM, these observables are constructed from the correlation function of two local chiral operators $\mathcal{O}_2(x)$, scalar operators of conformal dimension two belonging to the stress-tensor multiplet, in the presence of a line operator, such as a half-BPS Wilson line. By exploiting superconformal symmetry, one can show that this correlator is a nontrivial function of two conformal cross-ratios, as well as of the coupling constant \cite{Buchbinder:2012vr}. However, the dependence on the space--time coordinates can be eliminated by integrating over them with a measure that is completely fixed by superconformal invariance \cite{Dempsey:2024vkf,Billo:2023ncz,Billo:2024kri}. Moreover, the integrated correlator obtained by following this procedure can be computed using supersymmetric localization. Specifically, it is obtained by taking two derivatives of the vacuum expectation value of the Wilson loop, $\langle W \rangle$, in the mass-deformed $\mathcal{N}=2^{\star}$ theory obtained by giving a mass to the $\mathcal{N}=2$ hypermultiplet,
\begin{align}
\partial_{m}^2 \log \langle W \rangle \Big|_{m=0}
=
\int d^4x_1\, d^4x_2 \; \hat{\mu}(x_1,x_2)\,
\langle \mathcal{O}_2(x_1)\, \mathcal{O}_2(x_2) \rangle_{W}\, ,
\label{eq:IntegratedCorrelatorI}
\end{align}
where $\langle \mathcal{O}_2(x_1)\mathcal{O}_2(x_2) \rangle_{W}$ denotes the two-point function of the $\mathcal{O}_2(x)$ operators in the presence of the Wilson line, and $\hat{\mu}(x_1,x_2)$ is the integration measure mentioned above.

The relevance of these observables lies in the fact that they can be used to constrain bootstrap computations \cite{Chester:2022sqb}. Moreover, since line operators generally transform nontrivially under the $\mathcal{S}$-duality group \cite{MONTONEN1977117}, these correlators provide a powerful tool to investigate the implications of this symmetry. So far, they have been studied in the context of $\mathcal{N}=4$ SYM \cite{Dorigoni:2024vrb,Dorigoni:2024csx}, as well as in certain $\mathcal{N}=2$ SCFTs \cite{Pini:2024zwi,DeLillo:2025hal,DeSmet:2025mbc,DeLillo:2025eqg}. In the latter case, the operators $\mathcal{O}_2(x)$ are replaced by the so-called ``moment map operators'', namely scalar chiral operators of protected conformal dimension two belonging to the $\hat{\mathcal{B}}_1$ multiplet of the $\mathfrak{su}(2,2|2)$ superconformal algebra. Moreover, by considering the mass-deformed theory obtained by giving a mass to (some of) the $\mathcal{N}=2$ hypermultiplets, exact expressions can again be derived by exploiting supersymmetric localization, in a manner completely analogous to \eqref{eq:IntegratedCorrelatorI}

Among the various SCFTs analyzed so far, we believe it is natural to begin studying the resurgence properties of the integrated correlators \eqref{eq:IntegratedCorrelatorI} within the context of the $\mathcal{N}=4$ SYM theory with gauge group $Sp(N)$, as well as the $\mathcal{N}=2$ SCFT with matter content consisting of four hypermultiplets in the fundamental representation and one hypermultiplet in the rank-2 antisymmetric representation of $Sp(N)$ \cite{Beccaria:2021ism}. The motivation for this choice is threefold. First of all, these gauge theories admit a gravity dual geometry, which in the case of the $\mathcal{N}=2$ SCFT takes the form $AdS_5 \times S^5/\mathbb{Z}_2$ and is obtained through a suitable $\mathbb{Z}_2$ orientifold projection of the $\mathcal{N}=4$ SYM gravity dual \cite{Ennes:2000fu}. Therefore, the results obtained in this work constitute a natural starting point for a subsequent holographic analysis. Moreover, as it has been understood in \cite{Beccaria:2022kxy},  computations in these gauge theories can be efficiently performed using a Toda-like equation, which is a very powerful computational tool. Indeed, in the case of the $\mathcal{N}=2$ theory, it allows one to overcome the difficulties related to the non-Gaussianity of the corresponding matrix model, which is in general a non-trivial obstruction when attempting to obtain results in the large $N$ limit of a given $\mathcal{N}=2$ gauge theory. The use of this equation, in turn, permits the derivation of exact expressions, including the first non-perturbative corrections, for various observables such as the free energy and the vacuum expectation value of the 1/2 BPS Wilson loop \cite{Beccaria:2021ism,Beccaria:2022kxy}. Finally, in \cite{DeLillo:2025hal}, we obtained exact expressions for the integrated correlators with the insertion of a Wilson line in the large $N$ limit, for both gauge theories under consideration. These expressions can subsequently be expanded at strong coupling in the 't Hooft coupling $\lambda$, and thus provide the starting point for the resurgence analysis we aim to perform in this paper.

In particular, it is worth recalling that, while for the $\mathcal{N}=4$ SYM theory all the strong coupling expansions we found are given by infinite asymptotic series, in contrast, many observables of $\mathcal{N}=2$ SCFTs, which are needed to obtain the integrated correlators \eqref{eq:IntegratedCorrelatorI}, contain only a finite number of terms when expanded at strong coupling. Therefore, we would expect that the analysis of this latter case would require more attention and the use of the Cheshire cat resurgence techniques mentioned above. Although this is indeed possible, one further result of this article is to outline an alternative method to address these cases. As we will show, this approach is compatible with Cheshire cat resurgence, that is, it provides the same predictions for the instantonic corrections, while also having the advantage of yielding the results directly in terms of analytic functions, which can subsequently be easily expanded at strong coupling\footnote{In the present case, these will be modified Bessel $K_{\nu}(x)$ functions.}. Moreover, since it does not rely on the Toda-chain-like equation mentioned above, we expect that this method is not specific to the $Sp(N)$ theories considered in this article and, as we will argue, can be successfully applied to other $\mathcal{N}=2$ SCFTs and observables as well.

The rest of this article is organized as follows. In Section \ref{sec:MatrixModels}, we review how, by exploiting supersymmetric localization, a matrix model representation of the integrated correlators with the insertion of a Wilson line can be derived for both gauge theories with $Sp(N)$ gauge groups. Then, in Section \ref{Sec: Toda}, we review the main properties of the Toda-chain-like equation and how this computational tool can be employed to obtain exact expressions in the large $N$ limit and to derive the perturbative strong coupling expansions of the integrated correlators. After this, we are ready to perform the computation of the non-perturbative corrections. Specifically, in Section \ref{sec:N2theory}, we perform the computation for the $\mathcal{N}=2$ SCFT using a novel method complementary to Cheshire cat resurgence, while in Section \ref{sec:N4}, we perform the computation for the $\mathcal{N}=4$ SYM, where we also need to employ numerical techniques. Finally, in Section \ref{sec:Conclusions}, we draw our Conclusions. Our analysis is complemented by four appendices \ref{App:Cheschire}-\ref{app:PoleContributions} and an auxiliary \texttt{Mathematica} file, where we collect technical details concerning the derivation of the non-perturbative results and the exact expressions for the first three orders of the large-$N$ expansion of the integrated correlator for the $\mathcal{N}=2$ SCFT.

\section{The $Sp(N)$ theories and the corresponding matrix models}
\label{sec:MatrixModels}
We consider two distinct 4-dimensional gauge theories with gauge group $Sp(N)$. The first one is the $\mathcal{N}=4$ SYM theory, while the second is an $\mathcal{N}=2$ theory realized on a stack of $2N$ D3-branes, in the presence of eight D7-branes and a single O7 orientifold plane \cite{Aharony:1996en,Ennes:2000fu}. This gauge theory has four hypermultiplets in the fundamental representation and one hypermultiplet in the rank-2 antisymmetric representation of $Sp(N)$.

By placing the gauge theories under consideration on a unit 4-sphere $\mathbb{S}^4$ and applying supersymmetric localization \cite{Pestun:2007rz}, the computation of the corresponding partition function $\mathcal{Z}$ can be recast as
\begin{align}
\mathcal{Z} =  \int da\, \, \text{e}^{-\frac{8\pi^2N}{\lambda}\text{tr}\,a^2}\, \Big|\mathcal{Z}_{\text{1-loop}}(a)\,\mathcal{Z}_{\text{inst}}(a)\Big|^2\, \ ,
\label{Z}
\end{align}
where  $\lambda = g^2\,N$ is the 't Hooft coupling, while $a$ denotes a matrix belonging to the $\mathfrak{sp}(N)$ Lie algebra\footnote{
We employ the following conventions
\begin{align}
a = \sum_{b=1}^{N(2N+1)}a_bT^{b}\, ,\qquad \qquad  da = \prod_{b=1}^{N(2N+1)}\frac{d\,a_b}{\sqrt{2\pi}}\, \ ,    
\end{align}
where $T^{b}$ denote the generators of the $\mathfrak{sp}(N)$ Lie algebra in the fundamental representation, normalized such that $\text{tr}T^{a}T^{b}=\frac{1}{2}\delta^{a,b}$
.}. Finally $\mathcal{Z}_{\text{1-loop}}$ and $\mathcal{Z}_{\text{inst}}$ represent the 1-loop and instanton contributions respectively. In the large-$N$ limit we can neglect this last term, therefore we simply set  $\mathcal{Z}_{\text{inst}}(a)=1$. On the other hand, the 1-loop contribution depends on the  matter content of the theory.
As shown in \cite{Pufu:2023vwo,Billo:2023ncz,Dempsey:2024vkf}, in order to obtain the integrated correlator \eqref{eq:IntegratedCorrelatorI}, it is necessary to consider a mass deformation of the aforementioned theories. More specifically, in the case of $\mathcal{N}=4$ SYM we assign a mass $m$ to the adjoint hypermultiplet, thereby obtaining the so-called $\mathcal{N}=2^{\star}$ theory, while for the $\mathcal{N}=2$  theory we assign masses $m_f$ ($f=1,\ldots,4$) to the fundamental hypermultiplets. In both cases, one obtains a mass-dependent partition function, which needs to be expanded up to quadratic order in the mass. A detailed analysis of this computation was previously performed in \cite{DeLillo:2025hal}. Here, we simply report the results. In particular, after performing the rescaling 
\begin{align}
    a \mapsto \sqrt{\frac{\lambda}{8\pi^2N}}a \;,
    \label{ar}
\end{align}
the partition function for the $\mathcal{N}=2^{\star}$ theory reads
\begin{align}
\mathcal{Z}_{\mathcal{N}=2^{\star}}(m) =  \bigg(\frac{\lambda}{8\pi^2 N} \bigg)^{\frac{N(2N+1)}{2}} \int da \,\text{exp}\left[\displaystyle -\text{tr}a^2+m^2\,\widetilde{M}+O(m^4)\right]\, \ ,
\label{ZN2starExpansion}
\end{align}
where
\begin{align}
\widetilde{M} =  \widetilde{M}^{(1)} + \widetilde{M}^{(2)}\,.
\label{MN2star}
\end{align}
\label{SmallMassExpansionZSpData}
and the two contributions $\widetilde{M}^{(1)}$ and $\widetilde{M}^{(2)}$ are explicitly given by
\begin{subequations}
\begin{align}
& \widetilde{M}^{(1)} =  -\frac{1}{2}\sum_{n=1}^{\infty}(-1)^n(2n+1)\zeta_{2n+1}\left(\frac{\lambda}{2\pi^2N}\right)^{n}\text{tr}\,a^{2n} \, , \\[0.5em]
& \widetilde{M}^{(2)} = -\frac{1}{2}\sum_{n=1}^{\infty}\sum_{k=0}^{n}(-1)^n(2n+1)\zeta_{2n+1}\left(\frac{\lambda}{8\pi^2N}\right)^{n}\left(\begin{array}{c}
     2n  \\
     2k 
\end{array}\right) \text{tr}\,a^{2n-2k}\text{tr}\,a^{2k}\, \ ,
\end{align}
\label{M1andM2}
\end{subequations}
where $\zeta_{n}$ denotes  the value of the Riemann zeta function $\zeta(n)$.
Instead, for the mass-deformed $\mathcal{N}=2$ $Sp(N)$ theory we get 
\begin{align}
\mathcal{Z}(m_f) =  \, \Big(  \frac{\lambda}{8\pi^2N} \Big)^{\frac{N(2N+1)}{2}} 
\int da \,\text{exp}\bigg[-\text{tr}\,a^2-S_0+\sum_{f=1}^{4}m_f^2\,{M}+O(m_f^4)\,\bigg]\, \ ,
\label{Zmf}
\end{align}
where
\begin{align}
& S_0 = 4\sum_{n=1}^{\infty}(-1)^{n+1}\Big(\frac{\lambda}{8\pi^2N}\Big)^{n+1}(2^{2n}-1)\,\frac{\zeta_{2n+1}}{n+1}\,\text{tr}\,a^{2n+2}\,\  ,
\label{SOtilde}\\[0.5em]
& {M} = -\sum_{n=1}^{\infty}(-1)^n(2n+1)\,\zeta_{2n+1}\Big(\frac{\lambda}{8\pi^2N}\Big)^n\,\text{tr}\,a^{2n}\, \ .
\label{Msp}    
\end{align}
Finally, by recalling that the half-BPS circular Wilson loop in the fundamental representation admits a simple matrix-model description \cite{Pestun:2007rz, Beccaria:2021ism} in terms of the rescaled matrix $a$ \eqref{ar}, 
\begin{align}
W(a,\lambda) = 
\sum_{k=0}^{\infty}\frac{1}{k!}\Big(\frac{\lambda}{2N}\Big)^{\frac{k}{2}} \text{tr}a^{k} 
\label{WilsonLoopSp}   
\end{align}
and noting that its vacuum expectation values $\widetilde{\mathcal{W}}(m,\lambda)$ and $\mathcal{W}(m_f, \lambda)$ in the two mass-deformed theories can be straightforwardly computed, one obtains the matrix-model representation of the integrated correlator \eqref{eq:IntegratedCorrelatorI} for $\mathcal{N}=4$ SYM:
\begin{align}
& \widetilde{\mathcal{I}}(\lambda) \equiv \partial_{m}^2\log \widetilde{\mathcal{W}}(m,\lambda)\Big|_{m=0} = 2\,\frac{\langle W\,\widetilde{M} \rangle_{0} - \langle W \rangle_{0}\,\langle\,  \widetilde{M}\rangle_{0}}{\langle W \rangle_{0}} \, ,
\label{ISpN4}
\end{align}
where the subscript $_0$ indicates that the expectation values are computed in the Gaussian matrix model. Similarly, the matrix model representation for integrated correlator \eqref{eq:IntegratedCorrelatorI} of the $\mathcal{N}=2$ theory reads
\begin{align}
\mathcal{I}(\lambda) \equiv \partial_{m_f}^2\log \mathcal{W}(m_f,\lambda)\Big|_{m_f=0} = 2\frac{\langle W\,{M} \rangle - \langle W \rangle\,\langle {M}\rangle}{\langle W\rangle}\, \ ,
\label{ISpN2}    
\end{align}
where the v.e.v.'s in the right-hand side are computed in the  matrix model corresponding to the $\mathcal{N}=2$ theory.  For a generic function $f(a)$ such a v.e.v. is given by
\begin{align}
\langle f(a) \rangle = \frac{\langle f(a)\,\text{e}^{-S_0} \rangle_0}{\langle \text{e}^{-S_0} \rangle_0} \, \, ,
\end{align}
where $S_0$ is defined in \eqref{SOtilde}. 
Thus, once again, all computations reduce to evaluating Gaussian matrix-model integrals.

In the next Section, we briefly review how the large-$N$ expansions of the integrated correlators \eqref{ISpN4} and \eqref{ISpN2} can be efficiently obtained by exploiting a Toda-chain equation. The expressions obtained in this way are instrumental for the subsequent resurgence analysis.

\section{Toda chain and $Sp(N)$ SCFTs}
\label{Sec: Toda}
It is a well-known result that the partition function of a unitary matrix model with single-trace interaction potential satisfies a Toda-like integrable hierarchy \cite{ALVAREZGAUME199156,Morozov:2009uy,GERASIMOV1991565}. More recently, it has been shown that this property can be extended to symplectic gauge theories with single-trace potentials \cite{Beccaria:2021ism}. To make this article self-contained and to fix the notation for the following Sections, we briefly review these aspects here. In particular, the partition function of these theories satisfies the  equation
\begin{align}
\partial_{y}^2\log \mathcal{Z}_N(y) = \frac{\mathcal{Z}_{N+1}(y)\mathcal{Z}_{N-1}(y)}{\mathcal{Z}_N(y)^2}\, ,   
\label{eq:TodaChainZN}
\end{align}
supplemented with the boundary conditions
\begin{align}
\mathcal{Z}_{N=-1}(y) = 0\, , \quad \mathcal{Z}_{N=0}(y) = 1\, \ . 
\label{eq:boundary}
\end{align}
and with
\begin{align}
y = \frac{(4\pi)^2N}{\lambda}\, \ .   
\label{eq:Defy}
\end{align}
It then follows that the free energy $F_N = -\log \mathcal{Z}_N$ satisfies the equation
\begin{align}
\partial_{y}^2 F_N(y) = -\text{exp}\left[-F_{N+1}(y) +2F_N(y) -F_{N-1}(y)  \right]\,  . 
\label{eq:TodaChainFN}
\end{align}
As discussed in detail in \cite{Beccaria:2021ism,DeLillo:2025hal}, the equation \eqref{eq:TodaChainZN} allows one to derive a Toda chain equation governing a wide class of observables. Since this formalism will be extensively employed throughout the remainder of this work, we briefly review the main features of the Toda chain structure associated with the free energy and with the vacuum expectation values of the operators relevant for determining the large-$N$ expansion of the integrated correlator under consideration.
\subsection{Toda chain at large $N$}
Following \cite{Beccaria:2022kxy}, we adopt the following ansatz for the free energy of the $\mathcal{N}=2$ theory:
\begin{align}
F_N(\lambda) & \equiv F_N^{\mathcal{N}=4}(\lambda) + \Delta F_N(\lambda) \nonumber \\
&= F_N^{\mathcal{N}=4}(\lambda) + N F^{(1)}(\lambda) + F^{(2)}(\lambda) + \frac{F^{(3)}(\lambda)}{N} + O\left(\frac{1}{N^2}\right)\,,
\label{eq:FreeEnergyAtLargeN}
\end{align}
where $F_N^{\mathcal{N}=4}$ denotes the free energy of $\mathcal{N}=4$ SYM theory, given explicitly by \cite{Beccaria:2022kxy}
\begin{align}
F_N^{\mathcal{N}=4}(y) = \frac{N(2N+1)}{2}\log y - \log\left(\frac{G(N+1)\,G(N+3/2)}{G(3/2)}\right)\, ,
\label{eq:FreeEnergyN4}
\end{align}
where $G$ denotes the Barnes $G$-function. Apart from $F^{(1)}(\lambda)$, which, as shown in \cite{Beccaria:2021ism,DeLillo:2025hal}, admits the following integral representation:
\begin{align}
F^{(1)}(\lambda) = \frac{\log 2}{2\pi^2}\lambda + 4\int_0^{\infty}\frac{dt}{t}\frac{\text{e}^t}{(\text{e}^t+1)^2}\left[\left(\frac{2\pi}{\sqrt{\lambda}t}\right)J_1\left(\frac{t\sqrt{\lambda}}{\pi}\right)-1\right]\,,
\label{F1}
\end{align}
all other coefficients $F^{(i)}(\lambda)$ with $i \ge 2$ are determined by requiring that equation \eqref{eq:TodaChainFN} is satisfied and, importantly, are functions solely of $F^{(1)}(\lambda)$ and its derivatives. 

The next step is to derive the large-$N$ expansion of the vacuum expectation value of the Wilson loop, $\mathcal{W}_N$, the vacuum expectation value of the operator $M_N$ \eqref{Msp}, $\mathcal{M}_N$, and of the operator
\begin{align}
K_N = M_N\, W_N  - \mathcal{M}_N \mathcal{W}_N\, .
\label{KN}
\end{align}
Let us start by considering $\mathcal{M}_N$ and $\mathcal{W}_N$.  
At large $N$, we adopt the following ansatz:
\begin{subequations}
\begin{align}
\mathcal{M}_N &= N\,\mathcal{M}^{(0)}(\lambda) + \mathcal{M}^{(1)}(\lambda) + \frac{\mathcal{M}^{(2)}(\lambda)}{N} + O\left(\frac{1}{N^2}\right)\,,
\label{eq:MLargeN} \\
\mathcal{W}_N &= N\,\mathcal{W}^{(0)}(\lambda) + \mathcal{W}^{(1)}(\lambda) + \frac{\mathcal{W}^{(2)}(\lambda)}{N} + O\left(\frac{1}{N^2}\right)\,.
\label{eq:WLargeN}
\end{align}
\end{subequations}  

The case of the vacuum expectation value of the Wilson loop has been thoroughly analyzed in \cite{Beccaria:2021ism,Fiol:2014fla,ERICKSON2000155}, where it was shown that
\begin{align}
\mathcal{W}^{(0)}(\lambda) = \frac{4 I_1(\sqrt{\lambda})}{\sqrt{\lambda}}\,.
\label{W0}
\end{align}  
The leading coefficient $\mathcal{M}^{(0)}(\lambda)$ is not fixed by the Toda chain equation; however, as shown in \cite{DeLillo:2025hal}, it is given by
\begin{align}
\mathcal{M}^{(0)}(\lambda) = -2 \int_0^{\infty} dt\, \frac{\text{e}^t\,t}{(\text{e}^t-1)^2} \left[\left(\frac{4\pi}{\sqrt{\lambda}t}\right) J_1\left(\frac{\sqrt{\lambda}t}{2\pi}\right) - 1\right]\,.
\label{M0}
\end{align}  
Then, the Toda chain equation allows one to determine the first derivative of $\mathcal{M}^{(1)}$, which reads
\begin{align}
\partial_{\lambda}\mathcal{M}^{(1)}(\lambda) = -\frac{1}{4}\left(2\lambda^2 \partial_{\lambda}^2 F^{(1)}(\lambda) + 4\lambda \partial_{\lambda}F^{(1)}(\lambda) - 1 \right) \left(\lambda \partial_{\lambda}^2 \mathcal{M}^{(0)}(\lambda) + 2 \partial_{\lambda} \mathcal{M}^{(0)}(\lambda)\right)\,,
\label{DM1}
\end{align}  
and, all remaining coefficients $\mathcal{M}^{(j)}(\lambda)$ with $j \ge 2$ can be expressed solely in terms of $F^{(1)}(\lambda)$, $\mathcal{M}^{(0)}(\lambda)$, and their derivatives.\\
Finally, for the v.e.v. of the operator \eqref{KN}, $\mathcal{K}_N$, we adopt the following large-$N$ ansatz:
\begin{align}
\mathcal{K}_N(\lambda) = \mathcal{K}^{(0)}(\lambda) + \frac{\mathcal{K}^{(1)}(\lambda)}{N} + \frac{\mathcal{K}^{(2)}(\lambda)}{N^2} + O\left(\frac{1}{N^3}\right)\,,
\label{eq:KLargeN} 
\end{align}  
Using the Toda chain, one can show that the coefficients $\mathcal{K}^{(i)}(\lambda)$ with $i \ge 1$ are completely determined in terms of $F^{(1)}(\lambda)$, $\mathcal{M}^{(0)}(\lambda)$, $\mathcal{W}^{(0)}(\lambda)$, and their derivatives. By contrast, for $\mathcal{K}^{(0)}(\lambda)$ only its first derivative is fixed, and it is given by
\begin{align}
\partial_{\lambda}\mathcal{K}^{(0)}(\lambda) = \frac{1}{2\lambda} \left[ (\lambda^2 \partial_{\lambda}^2 + 2\lambda \partial_{\lambda}) \mathcal{W}^{(0)}(\lambda) \right] \left[ (\lambda^2 \partial_{\lambda}^2 + 2\lambda \partial_{\lambda}) \mathcal{M}^{(0)}(\lambda) \right]\,.
\label{DK0}
\end{align}  
However, by using the expressions \eqref{M0} and \eqref{W0}, and performing the integration over $\lambda$, it was found that \cite{DeLillo:2025hal}
\begin{align}
\mathcal{K}^{(0)}(\lambda) = 2\pi \int_0^{\infty} dt\, \frac{\text{e}^t\,t^2}{(1-\text{e}^t)^2} \frac{1}{4\pi^2+t^2}\, \mathcal{B}(t)\,,
\label{K0}
\end{align}  
where
\begin{align}
\mathcal{B}(t) \equiv \sqrt{\lambda}\, I_0(\sqrt{\lambda})\, J_1\left(\frac{t \sqrt{\lambda}}{2\pi}\right) - \frac{t \sqrt{\lambda}}{2\pi}\, I_1(\sqrt{\lambda})\, J_0\left(\frac{t \sqrt{\lambda}}{2\pi}\right)\,.
\label{B}
\end{align}

The large-$N$ results reviewed above will play a crucial role in the computation of the non-perturbative corrections to the integrated correlators with a Wilson line insertion, which will be analyzed in the following two Sections. We begin in the next Section by considering the $\mathcal{N}=2$ theory.

\section{Non-perturbative corrections in the  $\mathcal{N}=2$ $Sp(N)$ theory}
\label{sec:N2theory}
In this Section we derive the non-perturbative corrections to the strong-coupling expansion of the first three large-$N$ coefficients of the integrated correlator \eqref{ISpN2}, namely
\begin{align}
\mathcal{I}(\lambda) = \mathcal{I}^{(0)}(\lambda) + \frac{1}{N}\mathcal{I}^{(1)}(\lambda) + \frac{1}{N^2}\mathcal{I}^{(2)}(\lambda) + O\bigg(\frac{1}{N^3} \bigg)   \, \ ,
\end{align}
where the coefficients $\mathcal{I}^{(n)}$ can be computed by means of the recursion relation \cite{DeLillo:2025hal}
\begin{align}
\mathcal{I}^{(n)}(\lambda) =  \frac{2\,\mathcal{K}^{(n)}(\lambda)}{ \mathcal{W}^{(0)}(\lambda) } -\sum_{i=1}^{n}\frac{ \mathcal{W}^{(i)} (\lambda)}{\mathcal{W}^{(0)}(\lambda)}\,\mathcal{I}^{(n-i)}(\lambda)\, .
\label{IN2coefficients}
\end{align}
\subsection{An alternative route to non-perturbative corrections for truncated series}
Employing equation \eqref{IN2coefficients} we can extract the non-perturbative corrections to $\mathcal{I}^{(n)}$ at strong-coupling, provided
the full asymptotic expansions  of $\partial_{\lambda}F_1$, $\partial_{\lambda}\mathcal{M}^{(0)}$, $\mathcal{W}^{(0)}$ and $\mathcal{K}^{(0)}$ are known. We note that the non-perturbative contributions to $\mathcal{W}^{(0)}$ are completely encoded in \eqref{W0}, while the case of  $\partial_{\lambda}F_1$, $\partial_{\lambda}\mathcal{M}^{(0)}$ and $\mathcal{K}^{(0)}$ is much more challenging. We start by analyzing $\partial_{\lambda}F_1$ and $\partial_{\lambda}\mathcal{M}^{(0)} $, whose perturbative strong-coupling expansion has been obtained in Appendix A of \cite{DeLillo:2025hal} by applying the Mellin-Barnes technique and reads 
\begin{subequations}
\begin{align}
\partial_{\lambda}F^{(1)}(\lambda) &\underset{\lambda \rightarrow \infty}{\sim}  \frac{\log(2)}{2\pi^2} - \frac{1}{2\,\lambda} +  \frac{\pi^2}{2\lambda^2}\, 
\;, \label{StrongMF1}\\[1mm]
\partial_{\lambda}\mathcal{M}^{(0)}(\lambda) &\underset{\lambda \rightarrow \infty}{\sim}\, \frac{1}{\lambda} - \frac{4\pi^2}{3\lambda^2}\,\;,\\[1mm]
\mathcal{K}^{(0)}(\lambda) & \underset{\lambda \rightarrow \infty}{\sim} I_0(\sqrt{\lambda})\;.
\end{align}
\label{StrongM0F1}%
\end{subequations}
The crucial point is that the expansions \eqref{StrongM0F1} contain a finite number of terms and this fact prevents the use of standard resurgence techniques. This feature is also shared by $\mathcal{K}^{(0)}$, since in principle its large $\lambda$ expansion could include an infinite series of terms multiplying $I_0(\sqrt{\lambda})$, as well as a further contribution  proportional to $I_1(\sqrt{\lambda})$. However, these terms are absent because, when  applying Mellin Barnes techniques, the coefficient of the Bessel function $I_1(\sqrt{\lambda})$ turns out to be zero and the infinite series of terms multiplying $I_0(\sqrt{\lambda})$ collapses to a single term. Although it would be possible to exploit Cheshire Cat resurgence, which is specific to handle such cases, here we present an alternative strategy that leads to the resummation of our expressions in terms of modified Bessel $K_{\nu}$ functions of the second kind, which, at strong coupling, fully encodes the non-perturbative corrections.
For example, exploiting this method, we are able to rewrite $\partial_{\lambda}F^{(1)}$ in the exact-in-$\lambda$ form 
\begin{align}
\partial_{\lambda}F_1(\lambda) = \frac{\log(2)}{2\pi^2}-\frac{1}{2\lambda} + \frac{\pi^2}{2\lambda^2} +\frac{1}{\pi^2}\sum_{n=0}^{\infty}\left(K_0((2n+1)\sqrt{\lambda})-K_4((2n+1)\sqrt{\lambda})\right)\;.
\label{F1final}
\end{align}
This expression is compatible with equation  $(6.15)$ of \cite{Beccaria:2021ism}, that provides a resummation of $\partial_{\lambda}^2 (\lambda \widehat{F}^{(1)}(\lambda))$ in terms of Bessel $K_{\nu}$ functions, where $\widehat{F}^{(1)}$ is the planar component of the free energy of a $SU(N)$ theory such that $\widehat{F}^{(1)}= \frac{1}{2} F^{(1)}$. For completeness, in Appendix \ref{App:Cheschire} we show that the standard Cheschire Cat resurgence technique allows to recover its non-perturbative strong-coupling expansion but not the exact dependence in $\lambda$.
In order to show how our resummation strategy works, we start for simplicity with the example\footnote{We follow the same steps also in the case of a different index of the Bessel $J$ function and if the integration measure is 
\begin{align}
    \frac{\text{e}^t}{(\text{e}^t+1)^2} \;.
\end{align}} of $\partial_{\lambda}\mathcal{M}^{(0)}$ that, from \eqref{M0}, reads 
\begin{align}
\partial_{\lambda}\mathcal{M}^{(0)}(\lambda) = \frac{2}{\lambda} \int_{0}^{\infty}dt\;\frac{\text{e}^t\;t}{(\text{e}^t-1)^2} \; J_2 \bigg(\frac{t \sqrt{\lambda}}{\pi}  \bigg) \,.
\label{M0old}
\end{align}
At first we rewrite the integration measure by means of the identity 
\begin{align}
    \frac{\text{e}^t}{(\text{e}^t-1)^2} = \sum_{n=1}^{\infty} n\; \text{e}^{-nt} \;,
    \label{expid}
\end{align}
such that 
\begin{align}
&\partial_{\lambda}\mathcal{M}^{(0)}(\lambda)=  \frac{1}{2 \pi^2 a^2} \sum_{n=1}^{\infty} n \int_0^{\infty} dt\; t\; \text{e}^{-n t} J_2 \left( a t \right),
\end{align}
where $a=\sqrt{\lambda}/(2\pi)$. By using the series representation of the Bessel $J_2$ function and performing the integral over $t$, we get $\partial_{\lambda}\mathcal{M}^{(0)}(\lambda)= \mathcal{A}_1+\mathcal{A}_2$, where 
\begin{subequations}
\begin{align}
& \mathcal{A}_1= \frac{1}{ \pi ^2 a^4}\sum_{n=1}^{\infty} n \\
& \mathcal{A}_2= -\frac{1}{2 \pi ^2 a^2}\sum_{n=1}^{\infty}\bigg(\frac{n^2}{(a^2+n^2)^{3/2}} +\frac{2}{a^2}\frac{n^2}{(a^2+n^2)^{1/2}}\bigg) \;.
\label{A2}
\end{align}
\end{subequations}
$\mathcal{A}_1$ can be easily regularized exploiting the $\zeta$-function regularization, obtaining 
\begin{align}
    \mathcal{A}_1= -\frac{1}{12\pi^2 a^4},
    \label{A1}
\end{align}
while $\mathcal{A}_2$ requires more attention. In order to regularize it at first we use the identity 
\begin{align}
    \frac{1}{(n^2+a^2)^{\alpha}}= \frac{1}{\Gamma(\alpha)}\int_0^{\infty}dt\; t^{\alpha-1} \text{e}^{-(n^2+a^2)t}\;,
\end{align}
valid for $\alpha>0$\;, that allows to write
\begin{align}
    \mathcal{A}_2  = -\frac{1}{a^4\;\pi^{\frac{5}{2}}}\int_0^{\infty}dt\; \frac{1+a^2 t}{\sqrt{t}}\text{e}^{-a^2\;t} \sum_{n=1}^{\infty} n^2 \text{e}^{-n^2t}\;.
    \label{A2+A3}
\end{align}
\begin{comment}
$\mathcal{A}_2$ and $\mathcal{A}_3$ as 
\begin{subequations}
\begin{align}
    & \mathcal{A}_2 =  - \frac{1}{a^2\pi^\frac{5}{2}  \,} \int_0^{\infty} dt\; t^\frac{1}{2}\; e^{-a^2 t} \sum_{n=0}^{\infty} n^2 e^{-n^2 t } \;,\\
    & \mathcal{A}_3 =   -\frac{1}{a^4 \pi^{\frac{5}{2}}} \int_0^{\infty} dt\;t^{-\frac{1}{2}}\; e^{-a^2 t} \sum_{n=0}^{\infty} n^2 e^{-n^2 t}\; .
\end{align}
\label{A2A32}
\end{subequations}
\end{comment}
Now we exploit the Poisson summation formula in order to write 
\begin{align}
      \sum_{n=1}^{\infty} n^2 \text{e}^{-n^2 t} =   \frac{\sqrt{\pi}}{4 t^{\frac{3}{2}}} + \frac{\sqrt{\pi}}{4 t^{\frac{5}{2}}}\sum_{k=1}^{\infty} e^{-\frac{\pi^2 k^2}{t}}\; (2t -4 \pi^2 k^2)  \;,
 \label{PF}
\end{align}
getting 
\begin{align}
    \mathcal{A}_2= D - \frac{1}{4a^4\pi^2}\int_0^{\infty}dt\; \frac{1+a^2t}{t^3}\; \text{e}^{-a^2\;t} \sum_{k=1}^{\infty} (2t-4\pi^2k^2)\;\text{e}^{-\frac{\pi^2 k^2}{t}}\;,
    \label{A2PF}
\end{align}
where $D$ is the divergent integral 
\begin{align}
    D =  - \frac{1}{4a^2\pi^2} \int_0^{\infty}dt\; \frac{1+t}{t^2} e^{- t} \;.
\end{align}
In order to regularize it, we introduce the function
\begin{align}
    \mathcal{D}(z) \equiv  -\frac{1}{4a^2\pi^2}\int_0^{\infty}dt\; \frac{1+t}{t^2}\; t^z e^{-t} =- \frac{z}{4a^2\pi^2} \;\Gamma(z-1) \;, 
\end{align}
which is well defined for $\text{Re}(z)>1$ and reduces to $D$ for $z=0$. The right-hand-side, however, can be interpreted as the analytical continuation of the left-hand-side for $z \in \mathbb{C}$ except at $z = 1$ and $z = -n$, with $n=1,2,3,...$.  In particular it is analytic at $z=0$, yielding to
\begin{align}
    D= \mathcal{D}(0) = \frac{1}{4a^2\p^2}. 
\end{align}
Instead, the second term in \eqref{A2PF} can be expressed in terms of modified Bessel function of the second kind, exploiting the integral representation 
\begin{align}
    K_{\nu}(z) = \frac{1}{2}\bigg( \frac{z}{2}\bigg)^{\nu} \int_0^{\infty}dt \; t^{-\nu -1} \; e^{-t -\frac{z^2}{4t}}\;,
\end{align}
then we get 
\begin{align}
    \mathcal{A}_2= \frac{1}{4 \pi^2 a^2}  - \frac{2}{ a^2 \pi^2} \sum_{k=1}^{\infty} \left(\frac{K_1(2ak\pi)}{2ak\pi} -K_2(2ak \pi)  \right) - \frac{1}{a^2 \pi^2} \sum_{k=1}^{\infty}\left(K_0(2ak\pi)-2ak\pi K_1(2ak \pi)  \right)\;. \label{A2+A3Bessel}
\end{align}
Finally, summing \eqref{A1} and \eqref{A2+A3Bessel} and using the definition of $a$, after some algebra, we get
\begin{align}
    \partial_{\lambda} \mathcal{M}^{(0)}(\lambda)= \frac{1}{\lambda} -\frac{4\pi^2}{3\lambda^2} + \frac{4}{\lambda}\sum_{k=1}^{\infty}\left(K_0(\sqrt{\lambda}k) + \frac{(2+\lambda\,k^2)K_1(\sqrt{\lambda}k)}{\sqrt{\lambda}k}\right)\;.  
    \label{M0final}
\end{align}
The crucial point is that this representation of $\partial_{\lambda}\mathcal{M}^{(0)}$ in terms of Bessel $K_{\nu}$ functions is exact in $\lambda$  and we have verified numerically its compatibility with \eqref{M0old} for different values of the 't Hooft coupling. The great advantage of \eqref{M0final} is that the non-perturbative corrections are accessible simply by exploiting the asymptotic behavior of the Bessel $K$ functions for large values of their argument. 
Let's pass to $\mathcal{K}^{(0)}$. We start by expressing its derivative with respect to $\lambda$ in terms of Bessel $K_{\nu}$ functions by inserting \eqref{M0final} in \eqref{DK0}, obtaining 
\begin{align}
\partial_{\lambda}\mathcal{K}^{(0)}(\lambda) = \frac{I_1(\sqrt{\lambda})}{2\sqrt{\lambda}}\left(1+2\sqrt{\lambda}\sum_{k=1}^{\infty}\left(kK_1(\sqrt{\lambda}k)-k^2\sqrt{\lambda}K_0(\sqrt{\lambda}k)\right)\right)\, \ .
\label{DerivativeK0}
\end{align}
From the perturbative expansion of \eqref{K0} we know that $\mathcal{K}^{(0)}(0) = 0$, then we can write 
\begin{align}
\mathcal{K}^{(0)}(\lambda) = \int_0^{\lambda} dq\, \partial_{q}\mathcal{K}^{(0)}(q)
\label{K0integral} \;.
\end{align}
After having verified that the hypothesis of the Dominated Convergence Theorem are satisfied, as shown in Appendix \ref{App:DominatedConvergence}, we exchange the integral and the infinite sum in \eqref{K0integral} and we perform the integral term by term, getting
\begin{align}
\mathcal{K}^{(0)}(\lambda)= I_0(\sqrt{\lambda})
   + \sum_{k=1}^{\infty} \bigl( \mathcal{J}_0(k,\lambda) + \mathcal{J}_1(k,\lambda) \bigr)
   -  \frac{7}{4} - \frac{\pi^2}{3} \, ,
\label{K0final}
\end{align}
where \begin{align}
\mathcal{J}_0(k,\lambda) =  
\begin{cases}
\displaystyle
-\frac{1}{2} \lambda^{3/2} 
\Big( I_1(\sqrt{\lambda}) K_0(\sqrt{\lambda}) 
+ I_2(\sqrt{\lambda}) K_1(\sqrt{\lambda}) \Big), & k=1, \\[1em]
\displaystyle
-\frac{2 \lambda k^2 I_0(\sqrt{\lambda}) K_0(k \sqrt{\lambda})}{(k^2-1)^2} 
+ \frac{4 \sqrt{\lambda} k^2 I_1(\sqrt{\lambda}) K_0(k \sqrt{\lambda})}{(k^2-1)^2} 
+ \frac{2 \lambda k^5 I_1(\sqrt{\lambda}) K_1(k \sqrt{\lambda})}{(k^2-1)^2} \\[0.5em]
\displaystyle
+ \frac{2 \lambda k^4 I_0(\sqrt{\lambda}) K_0(k \sqrt{\lambda})}{(k^2-1)^2} 
+ \frac{4 \sqrt{\lambda} k^3 I_0(\sqrt{\lambda}) K_1(k \sqrt{\lambda})}{(k^2-1)^2} 
- \frac{2 \lambda k^3 I_1(\sqrt{\lambda}) K_1(k \sqrt{\lambda})}{(k^2-1)^2}, & k \ge 2
\end{cases}
\label{integralK0I}
\end{align}
and
\begin{align}
\mathcal{J}_1(k,\lambda) 
=
\begin{cases}
\displaystyle
\lambda I_1(\sqrt{\lambda}) K_1(\sqrt{\lambda}) 
+ \lambda I_0(\sqrt{\lambda}) \left(K_0(\sqrt{\lambda}) +\frac{2}{z}K_1(\sqrt{\lambda})\right), & k=1, \\[0.5em]
\displaystyle
- \frac{2 k \sqrt{\lambda} \Big( k I_1(\sqrt{\lambda}) K_0(k \sqrt{\lambda}) 
+ I_0(\sqrt{\lambda}) K_1(k \sqrt{\lambda}) \Big)}{k^2 - 1}, & k \ge 2\;.
\end{cases}
\label{integralK1I}
\end{align}
We stress that equation \eqref{K0final} is exact in $\lambda$.
Exploiting \eqref{F1final}, \eqref{M0final} and \eqref{K0final} as well as the asymptotic behavior of the Bessel $I$ and $K$ functions, we are able to obtain the non-perturbative corrections at large $\lambda$ to all the large-$N$ coefficients $\mathcal{K}^{(i)}$, with $i \ge 1$, and $\mathcal{W}^{(j)}$, with $j\ge 2$, although the computation becomes increasingly involved. The only remaining case that requires more attention is $\mathcal{W}^{(1)}$, because, as shown in \cite{Beccaria:2022kxy} and reviewed in Section \ref{Sec: Toda}, the Toda Chain equation fixes only its first derivative and not the function itself. Then, using the analog of equation \eqref{DM1} and following the same strategy adopted for $\mathcal{K}^{(0)}$, we get 
\begin{align}
\mathcal{W}^{(1)}(\lambda) 
&= I_0(\sqrt{\lambda}) 
   - \lambda \frac{\log(2)}{\pi^2} I_2(\sqrt{\lambda}) 
   +  \frac{1}{2\pi^2} - \frac{2}{3}  \nonumber\\
&\quad - \frac{2}{\pi^2} \sum_{k=0}^{\infty} \frac{1}{(2k+1)^2} 
      \bigl( \mathcal{J}_1(2k+1,\lambda) - \mathcal{J}_0(2k+1,\lambda) \bigr),
\label{W1}      
\end{align}
where $\mathcal{J}_0$ and $\mathcal{J}_1$ are the same quantities given in \eqref{integralK0I} and \eqref{integralK1I}. 
\subsection{The non-perturbative strong coupling expansions of $\mathcal{I}$}
Now we are finally able to determine the exponentially small corrections to the strong coupling expansions of the coefficients $\mathcal{I}^{(n)}$. We start from the planar term $\mathcal{I}^{(0)}$, that, from \eqref{IN2coefficients}, \eqref{W0} and \eqref{K0final}, after some algebra reads 
\begin{align}
\mathcal{I}^{(0)}(\lambda)
&=
\frac{\sqrt{\lambda}}{2}\,\frac{I_0(\sqrt{\lambda})}{I_1(\sqrt{\lambda})}
-
\frac{\sqrt{\lambda}}{2 I_1(\sqrt{\lambda})}
\left(\frac{7}{4}+\frac{\pi^2}{3}\right)
+\lambda^{3/2} K_1(\sqrt{\lambda})
-\frac{\lambda^2}{4} K_0(\sqrt{\lambda})
\nonumber \\[4pt]
&\quad
-\frac{I_0(\sqrt{\lambda})}{4 I_1(\sqrt{\lambda})}
\!\left[
(\lambda-4)\lambda\, K_1(\sqrt{\lambda})
-2\lambda^{3/2} K_0(\sqrt{\lambda})
\right]
\nonumber \\[6pt]
&\quad
+\lambda
\sum_{k=2}^{\infty}
\frac{k}{(k^2-1)^2}
\Bigg[
\frac{I_0(\sqrt{\lambda})}{I_1(\sqrt{\lambda})}
\,A_k(\lambda)
+ k\,B_k(\lambda)
\Bigg],
\label{I0exact}
\end{align}
where
\begin{subequations}
\begin{align}
A_k(\lambda)
&=
k(k^2-1)\sqrt{\lambda}\,K_0(k\sqrt{\lambda})
+(k^2+1)K_1(k\sqrt{\lambda}),
\\[4pt]
B_k(\lambda)
&=
(3-k^2)\,K_0(k\sqrt{\lambda})
+k(k^2-1)\sqrt{\lambda}\,K_1(k\sqrt{\lambda}).
\end{align}
\end{subequations}
By exploiting the properties of the Bessel functions at strong coupling, we can readily obtain
\begin{align}
\mathcal{I}^{(0)}(\lambda)
\underset{\lambda \to \infty}{\sim}\;
&\frac{\sqrt{\lambda}}{2}
 + \frac{1}{4}
 + \frac{3}{16\,\sqrt{\lambda}}
 + \frac{3}{16\,\lambda}
 + \frac{63}{256\,\lambda^{3/2}}
 + O\!\left(\frac{1}{\lambda^{2}}\right)
\nonumber \\[6pt]
&\quad
+ \text{e}^{-\sqrt{\lambda}}
  \sqrt{\frac{\pi}{2}}\,
  \Bigg[
      -\frac{\lambda^{7/4}}{2}
      + \frac{21}{16}\,\lambda^{5/4}
      - \frac{243 + 256\pi^{2}}{768}\,\lambda^{3/4}
      + \frac{351 - 256\pi^{2}}{2048}\,\lambda^{1/4}
      +  O\!\left(\frac{1}{\lambda^{1/4}}\right)
  \Bigg]
\nonumber \\[8pt]
&\quad
+ \text{e}^{-2\sqrt{\lambda}}
  \Bigg[
      2\sqrt{\pi}\,\lambda^{5/4}
      + \frac{7}{8}\sqrt{\pi}\,\lambda^{3/4}
      + i\,\sqrt{\lambda}
      + \frac{153}{256}\sqrt{\pi}\,\lambda^{1/4}
      + \frac{3i}{4}
      + O\!\left(\frac{1}{\lambda^{1/4}}\right)
  \Bigg]
\nonumber \\[6pt]
&\quad
+ O\!\left(\text{e}^{-3\sqrt{\lambda}}\right)\, .
\label{StrongI0}
\end{align}
Following the same procedure we explicitly determine the non-perturbative corrections to the strong coupling expansions of $\mathcal{I}^{(1)}$ and $\mathcal{I}^{(2)}$. Here we reports only the first orders of the expansions but more details can be found in the ancillary \texttt{Mathematica} file. We get
\begin{subequations}
\begin{align}
\mathcal{I}^{(1)}(\lambda) \underset{\lambda\to\infty}{\sim}\;&
-\frac{\lambda^{3/2}\log 2}{8\pi^{2}}
-\frac{\sqrt{\lambda}}{8}
+\frac{3\sqrt{\lambda}\log 2}{64\pi^{2}}
-\frac{1}{8}
+\frac{3\log 2}{32\pi^{2}}
+O(\lambda^{-1/2}) \nonumber
\\[8pt]
&\quad
+\text{e}^{-\sqrt{\lambda}}\Bigg[
-\frac{\lambda^{13/4}\log 2}{8\sqrt{2}\,\pi^{3/2}}
+\frac{49\,\lambda^{11/4}\log 2}{64\sqrt{2}\,\pi^{3/2}} \nonumber
\\
&\qquad\qquad
+\left(
\frac{1}{4\sqrt{2}\pi^{3/2}}
+\frac{1}{8}\sqrt{\frac{\pi}{2}}
-\frac{921\log 2}{1024\sqrt{2}\pi^{3/2}}
-\frac{1}{12}\sqrt{\frac{\pi}{2}}\log 2
\right)\lambda^{9/4}
+O(\lambda^{7/4})
\Bigg] \nonumber
\\[8pt]
&\quad
+\text{e}^{-2\sqrt{\lambda}}\Bigg[
-\frac{\lambda^{7/2}}{8\pi}
+\frac{5\lambda^{3}}{32\pi}
+\frac{\lambda^{11/4}\log 2}{\pi^{3/2}} \nonumber
\\
&\qquad\qquad
-\frac{13\lambda^{9/4}\log 2}{16\pi^{3/2}}
+\left(
\frac{155}{256\pi}
-\frac{\pi}{6}
\right)\lambda^{5/2}
+O(\lambda^{2})
\Bigg] + O(\text{e}^{-3\sqrt{\lambda}})\;, \label{I1} \\
\mathcal{I}^{(2)} (\lambda)\underset{\lambda \to \infty}{\sim}\;&
\frac{3\log^{2}2}{64\pi^{4}}\,\lambda^{5/2}
+ \left(
 -\frac{1}{768}
 - \frac{3\log^{2}2}{512\pi^{4}}
 + \frac{\log 256}{256\pi^{2}}
 \right)\lambda^{3/2}
+ \frac{3}{256}\,\lambda
+ O(\sqrt{\lambda})
\nonumber \\[6pt]
&\quad
+ \text{e}^{-\sqrt{\lambda}}\!\left[
 -\frac{\log^{2}2}{64\sqrt{2}\,\pi^{7/2}}\,\lambda^{19/4}
 + \frac{101\log^{2}2}{512\sqrt{2}\,\pi^{7/2}}\,\lambda^{17/4}
 + \left(
 -\frac{5233\log^{2}2}{8192\sqrt{2}\,\pi^{7/2}}
 -\frac{\log^{2}2}{96\sqrt{2}\,\pi^{3/2}}
 \right.\right.
\nonumber \\[3pt]
&\qquad\qquad\qquad\left.\left.
 + \frac{\log 2}{16\sqrt{2}\,\pi^{7/2}}
 + \frac{\log 2}{32\sqrt{2}\,\pi^{3/2}}
 \right)\lambda^{15/4}
 + O(\lambda^{13/4})
\right]
\nonumber \\[8pt]
&\quad
+ \text{e}^{-2\sqrt{\lambda}}\!\left[
 -\frac{\log 2}{16\pi^{3}}\,\lambda^{5}
 + \frac{19\log 2}{64\pi^{3}}\,\lambda^{9/2}
 + \frac{\log^{2}2}{4\pi^{7/2}}\,\lambda^{17/4}
 + O(\lambda^{4})
\right]
\nonumber \\[6pt]
&\quad + O(\text{e}^{-3\sqrt{\lambda}})\;.
\label{I2}
\end{align}
\label{I1I2}
\end{subequations}
It is worth noting that the first lines of \eqref{StrongI0}, \eqref{I1} and \eqref{I2} encode the perturbative sector of these strong-coupling expansions and coincide with those reported in \cite{DeLillo:2025hal}. 

\section{Non-perturbative corrections in the $\mathcal{N}=4$ SYM theory}
\label{sec:N4}
In this Section we compute the leading non-perturbative planar and next-to-planar contributions to the large-$N$ expansion of the integrated correlator of the $\mathcal{N}=4$ SYM theory \eqref{ISpN4}, namely
\begin{align}
\widetilde{\mathcal{I}}(\lambda) \, = \,
\widetilde{\mathcal{I}}^{(0)}(\lambda)
+
\frac{1}{N}\,\widetilde{\mathcal{I}}^{(1)}(\lambda)
+
\mathcal{O}\!\left(\frac{1}{N^{2}}\right)\,.
\label{eq:IN4LargeN}
\end{align}

We begin by recalling the results of \cite{DeLillo:2025hal}, where the planar term was shown to admit the integral representation
\begin{align}
\widetilde{\mathcal{I}}^{(0)}(\lambda)
=
\frac{8\pi^{2}}{I_{1}(\sqrt{\lambda})}
\int_{0}^{\infty} \! dt\,
\frac{\text{e}^{t}\, t}{(\text{e}^{t}-1)^{2}}\,
\frac{1}{4\pi^{2}+t^{2}}\,
J_{1}\!\left(\frac{\sqrt{\lambda}\, t}{2\pi}\right)
\mathcal{B}(t)\,,
\label{I0tilde}
\end{align}
with $\mathcal{B}(t)$ defined in \eqref{B}. The next-to-planar contribution can be written  as
\begin{align}
\widetilde{\mathcal{I}}^{(1)}(\lambda)
=
\frac{1}{\mathcal{W}^{(0)}(\lambda)}
\left[
2\,\widetilde{\mathcal{K}}^{(1)}(\lambda)
-
\frac{1}{2}
\bigl(I_{0}(\sqrt{\lambda})-1\bigr)\,
\widetilde{\mathcal{I}}^{(0)}(\lambda)
\right],
\label{I1tilde}
\end{align}
where $\mathcal{W}^{(0)}$ is given in \eqref{W0} and the explicit expression for $\widetilde{\mathcal{K}}^{(1)}$ reads
\begin{align}
\widetilde{\mathcal{K}}^{(1)}(\lambda)
&=
\frac{\pi}{2}
\int_{0}^{\infty} \! dt\,
\frac{\text{e}^{t}\, t^{2}}{(1-\text{e}^{t})^{2}}\,
\frac{1}{\pi^{2}+t^{2}}\,
\mathcal{B}(2t)
\nonumber\\
&\quad
+ \frac{\sqrt{\lambda}}{2}\,
I_{1}(\sqrt{\lambda})
\int_{0}^{\infty} \! dt\,
\frac{\text{e}^{t}\, t}{(1-\text{e}^{t})^{2}}\,
J_{1}\!\left(\frac{t\sqrt{\lambda}}{2\pi}\right)^{2}
\nonumber\\
&\quad
- \pi
\int_{0}^{\infty} \! dt\,
\frac{\text{e}^{t}\, t^{2}}{(1-\text{e}^{t})^{2}}\,
\frac{1}{4\pi^{2}+t^{2}}\,
\Bigl[
1 - J_{0}\!\left(\frac{t\sqrt{\lambda}}{2\pi}\right)
\Bigr]
\mathcal{B}(t)\, .
\label{K1tilde}
\end{align}
In the following we analyze separately the planar and next-to-planar contributions \eqref{I0tilde} and \eqref{I1tilde}, starting with the planar term. In doing so, we also review the numerical techniques that will be employed in both cases and that are needed, since the presence of a different integration measure and integrands quadratic in the Bessel $J$ functions prevents us to employ the method of Section \ref{sec:N2theory}.
\subsection{The planar term}
\label{subsec:PlanarTerm}
\begin{figure}[t]
    \centering
\includegraphics[width=0.9\textwidth]{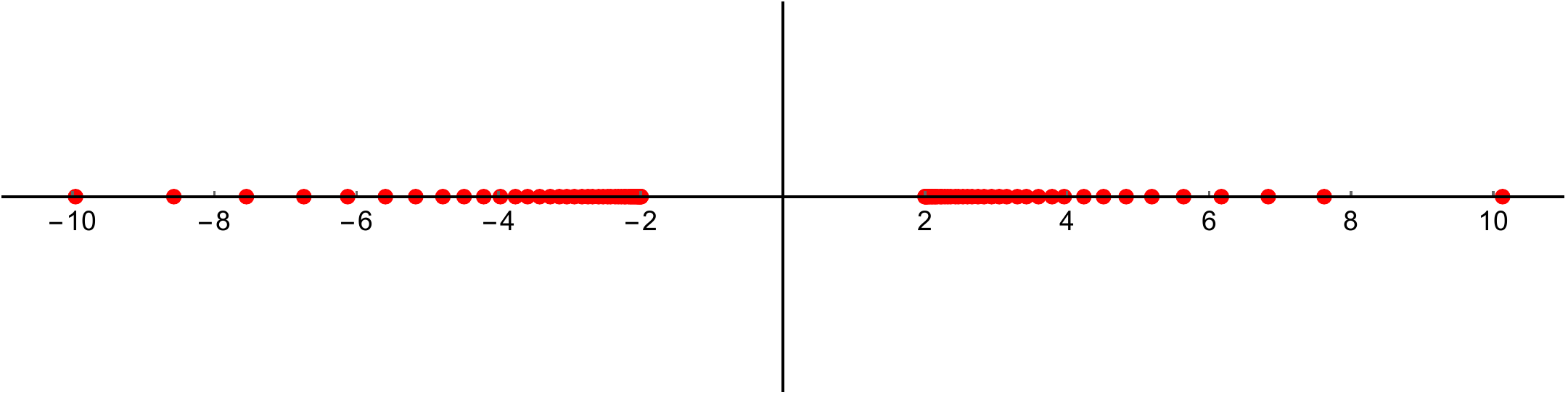}
    \caption{Poles of the diagonal Borel-Padé approximant $PB_{[100/100]}[\widetilde{\mathcal{I}}^{(0)}/\sqrt{\lambda}]$. The poles are in $t$-variable after the transformation \eqref{BorelI0}.}
\label{fig:PadeI0tilde}
\end{figure}

\begin{figure}[t]
  \centering
  \begin{subfigure}{0.48\textwidth}
    \centering
\includegraphics[width=\linewidth]{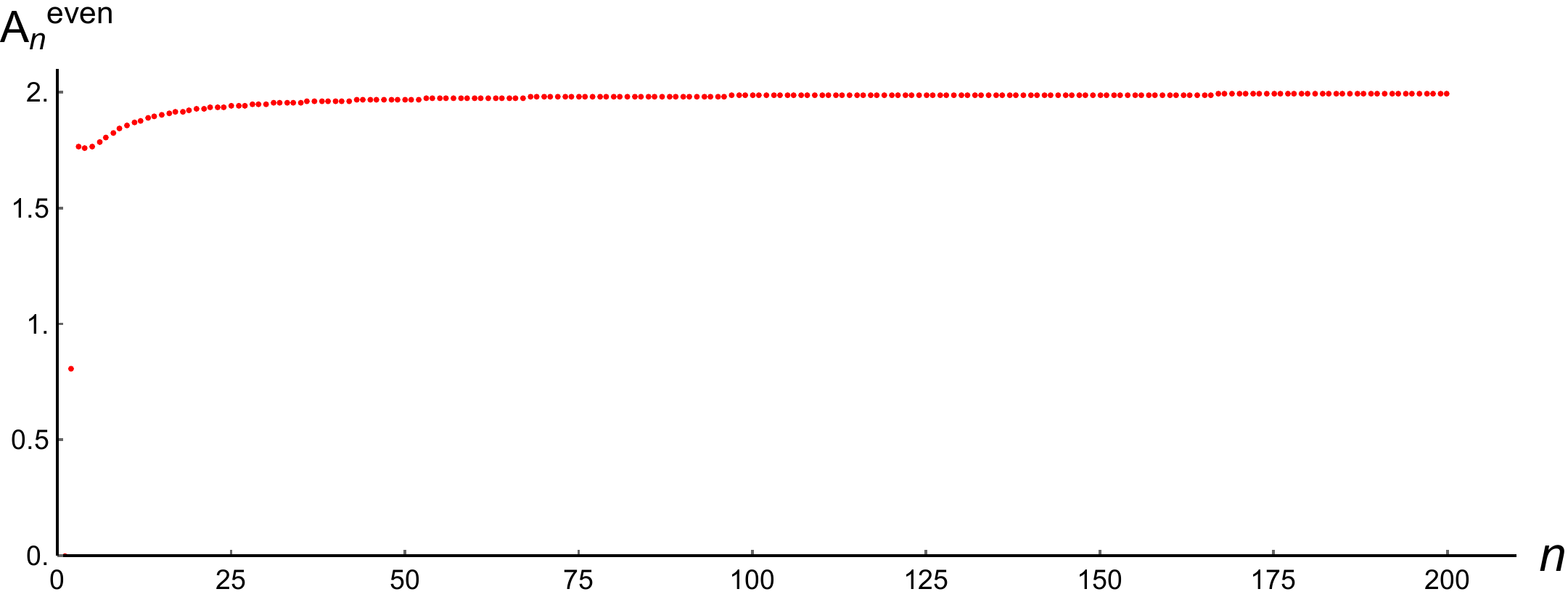}
  \end{subfigure}\hfill
  \begin{subfigure}{0.48\textwidth}
    \centering    \includegraphics[width=\linewidth]{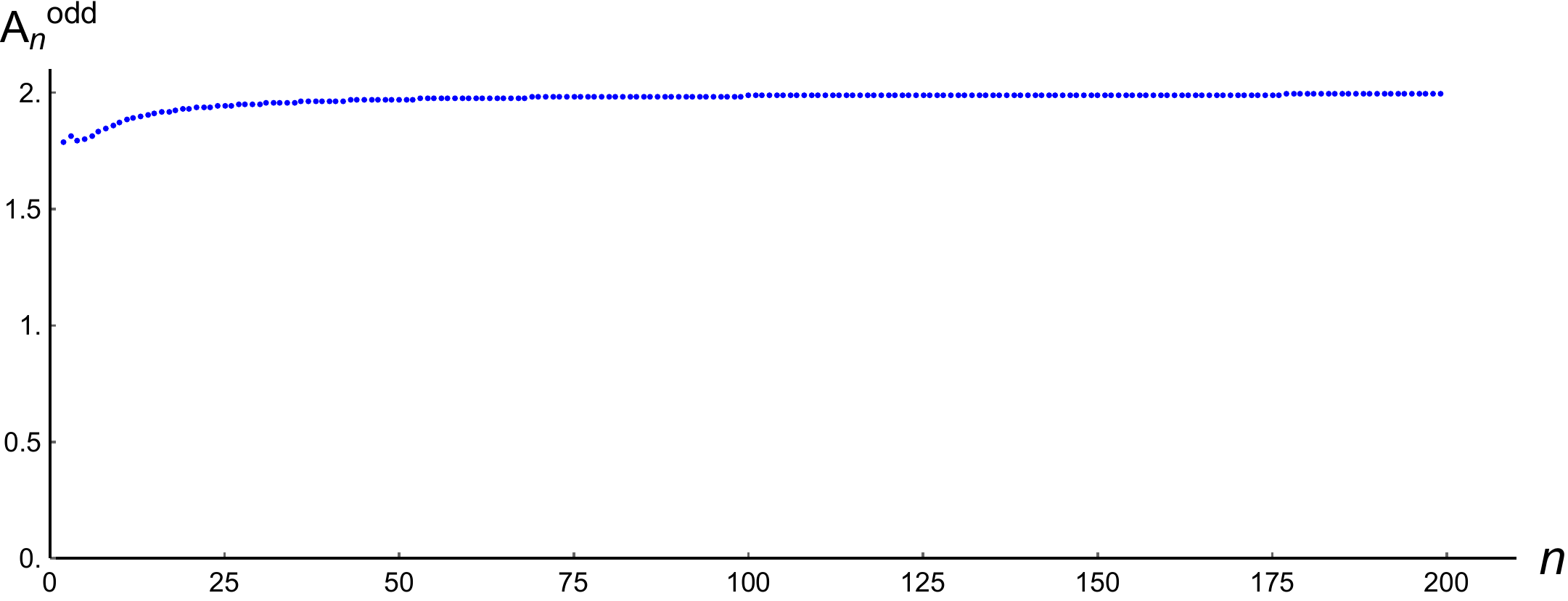}
  \end{subfigure}
  \caption{The  sequence $A_n^{\text{even}}$ defined in \eqref{eq:AEvenModule} is shown on the left, while the sequence $A_n^{\text{odd}}$ \eqref{eq:AOddModule} constructed from the odd coefficients  is shown on the right. In both cases, the numerical data seem to converge very rapidly to the value $2$.}
  \label{fig:AEvenOdd}
\end{figure}

Using the Mellin–Barnes representation of the Bessel function, we find that the planar contribution~\eqref{I0tilde} admits the following large-$\lambda$ expansion:
\begin{align}
\widetilde{\mathcal{I}}^{(0)}(\lambda) \;\underset{\lambda \to \infty}{\sim}\; \mathcal{Q}_{(1,0)}(\lambda) + \frac{I_0(\sqrt{\lambda})}{I_1(\sqrt{\lambda})}\,\mathcal{Q}_{(1,1)}(\lambda) \, \equiv \, \sum_{n=0}^{\infty}a_n\lambda^{\frac{1-n}{2}}\, \ ,
\label{I0strong}
\end{align}
where it is understood that the ratio of Bessel functions is expanded at strong coupling. The labels of the two coefficients, $\mathcal{Q}_{(1,0)}$ and $\mathcal{Q}_{(1,1)}$, are chosen to indicate that these terms arise from the contributions in \eqref{I0tilde} containing, respectively, one $J_0$ and one $J_1$, and two $J_1$ Bessel functions. Using the techniques developed in Appendix B of \cite{Pufu:2023vwo}, after a lengthy but straightforward computation, we find
\begin{subequations}
\begin{align}
& \mathcal{Q}_{(1,0)}(\lambda) =  -1 +\frac{1}{3}\left(\frac{\lambda}{\pi}\right)^{3/2}\sum _{s=1}^{\infty }\frac{1}{\lambda^{s}}\frac{\Gamma\left(s-\frac{3}{2}\right)^2 \Gamma\left(s-\frac{1}{2}\right)}{\Gamma(s-1)}\left(12 \sum _{i=1}^{s-2} i \zeta_{2i+1}+\pi ^2-3\right)\, \ ,
\label{Q10}
\\[0.5em]
&  \mathcal{Q}_{(1,1)}(\lambda) = \,   \,  \sqrt{\lambda } + \frac{\lambda^2}{3\pi^{3/2}}\sum _{s=1}^{\infty }\frac{1}{\lambda^{s}} \frac{\Gamma\left(s-\frac{5}{2}\right)\Gamma\left(s-\frac{3}{2}\right)\Gamma\left(s-\frac{1}{2}\right) }{\Gamma(s-1)}\left(12 \sum _{i=1}^{s-2}i\zeta_{2i+1}+\pi ^2-3\right)\, \ ,
\label{Q11}
\end{align}
\end{subequations}
and, therefore, the $a_n$ coefficients appearing in the r.h.s. of \eqref{I0strong} can be easily determined. We observe that, in contrast to the situation encountered in Section \ref{sec:N2theory}, the large-$\lambda$ expansion of \eqref{I0tilde} contains an infinite number of terms. Therefore, in order to compute the leading non-perturbative corrections, our analysis will rely on resurgence techniques reviewed in \cite{Dorigoni:2014hea}. To this end, the first step is to consider the Borel transform $\mathbf{B}$ associated with the expansion \eqref{I0strong}, namely
\begin{align}
\sum_{n\geq 0} a_n\, z^{n} \;\longmapsto\;
\mathbf{B}\!\left[\frac{1}{\sqrt{\lambda}}\,\widetilde{\mathcal{I}}^{(0)}\right]
\equiv \sum_{n\geq 0} a_n \frac{t^n}{n!}\, ,
\label{BorelI0}
\end{align}
where, for notational convenience, we have set $z=\lambda^{-1/2}$. Moreover, we have multiplied $\widetilde{\mathcal{I}}^{(0)}$ by  $\lambda^{-1/2}$ so that the resulting series in $z$ contains only non-negative powers.  It's not straightforward to determine whether the Borel transform \eqref{BorelI0} admits a closed-form expression and we therefore construct its diagonal Borel--Pad\'e approximant of order $M$, denoted by $BP_{[M/M]}[\widetilde{\mathcal{I}}^{(0)}/\sqrt{\lambda}]$, which is given by the ratio of two polynomials of degree $M$. As it is well established \cite{Costin:2021bay,Dunne:2025mye}, although the Borel--Pad\'e approximant is obtained through a numerical procedure, it encodes a substantial amount of information about the analytic structure of the Borel transform. In particular, the singularity structure of the latter can be inferred from the distribution of poles of the Borel--Pad\'e approximant.  We construct the Borel--Pad\'e approximant for several increasing values of $M$, up to $M=100$. Importantly, for sufficiently large $M$, our results become insensitive to the degree of the approximating polynomials, thereby providing strong evidence for the robustness of our analysis. In Figure \ref{fig:PadeI0tilde} we report the pole structure of the diagonal Borel–Padé approximant obtained for $M=100$.
From this analysis, it is evident that $t = \pm 2$ are accumulation points of the poles.  We therefore expect the Borel transform associated with the expansion \eqref{I0strong} to be analytic in a neighborhood of $t=0$, with its leading singularities located on the circle of radius $2$ centered at the origin. Moreover, the observed accumulation of poles strongly indicates the presence of two branch cuts, originating at $t=\pm 2$ and extending along the real axis towards $\pm\infty$, respectively. Then, following \cite{Dorigoni:2014hea,marino_resurgence_course} and further assuming that the singularities of the Borel transform are either poles, logarithmic branch cuts, or branch cuts of the form
\begin{align}
(t - t_{\star})^{-b} \qquad \text{with} \qquad 0 < b < 1 \, \ ,
\end{align}
we are led to consider the following ansatz for the large-$n$ behavior of the coefficients $a_n$, which accounts for the presence of two singularities\footnote{Although the previous numerical analysis strongly suggests that $|A|=2$, we chose to remain agnostic about the locations of the singularities. As a consistency check, we will subsequently recover the expected value of $|A|$.} located at the generic points $\pm A$:
\begin{equation}
\begin{aligned}
a_n  \, \underset{n>>1}{\sim} \, & \,
\frac{\mathcal{S}_1}{2\pi i} \Biggl( 
\frac{f(b_1)}{A^{\,n+b_1}} \sum_{m\ge 0} A^m c_m^{(1)} \Gamma(n+b_1-m) 
+ \frac{\Gamma(n+1)}{A^{\,n+b_1+1}} 
\Biggr) \\
& + \frac{\mathcal{S}_2}{2\pi i} \Biggl( 
\frac{f(b_2)}{(-A)^{\,n+b_2}} \sum_{m\ge 0} (-A)^m c_m^{(2)} \Gamma(n+b_2-m) 
+ \frac{\Gamma(n+1)}{(-A)^{\,n+b_2+1}} 
\Biggr) \, .
\end{aligned}
\label{anLarge}
\end{equation}
where $\mathcal{S}_1$ and $\mathcal{S}_2$ are the Stokes constants associated with the two singularities, while the function $f(b)$ depends on the type of branch cut and is defined as \cite{marino_resurgence_course}
\begin{align}
f(b) =
\begin{cases} 
\sin(\pi b) & 0 < b < 1, \\[1mm]
1 & b = 0,
\end{cases}
\end{align}
the latter corresponding to a logarithmic branch cut. Our goal is to determine the first coefficients $c_n^{(1)}$ together with the Stokes constant $\mathcal{S}_1$, which encode the first non-perturbative contribution to \eqref{I0tilde}. To this end, as a first step, we determine the parameters $A$, $b_1$ and $b_2$ appearing in \eqref{anLarge}. 
To simplify the discussion and illustrate the method, let us focus on the determination of $A$. Starting from \eqref{anLarge} and after some algebra, we find that the convergence radius of \eqref{BorelI0} can be numerically evaluated by considering the following sequences and estimating their limiting values
\begin{subequations}
\begin{align}
A_n^{\text{even}}  & \equiv n \, \sqrt{\frac{a_n}{a_{n+2}}} \, \qquad \text{with} \qquad \, n = 0, 2, 4, 6, \ldots \,  ,
\label{eq:AEvenModule} \\
A_n^{\text{odd}} & \equiv n \, \sqrt{\frac{a_n}{a_{n+2}}} \, \qquad \text{with} \qquad \, n = 1, 3, 5, 7, \ldots \,  .
\label{eq:AOddModule}
\end{align}
\label{eq:Amodule}
\end{subequations}These two sequences, up to $n=200$, are shown in Figure \ref{fig:AEvenOdd}.
 For instance, for $n=150$ we obtain
\begin{align}
A_{150}^{\text{even}} = 1.990020\ldots \, , \qquad A_{150}^{\text{odd}} =  1.990097\ldots \, \ .
\label{AsequenceValues}
\end{align}
These results strongly indicate that both sequences converge to $|A| = 2$. This estimate can be further improved by recalling that, for a generic sequence $s_n$ with an asymptotic expansion of the form
\begin{align}
s_n = \sum_{k=0}^{\infty} \frac{g_k}{n^k} \,,
\label{sn}
\end{align}
one can define its $M$-th Richardson transform as
\begin{align}
s_n^{(M)} = \sum_{\ell=0}^{M} \frac{s_{n+\ell} \, (n+\ell)^M \, (-1)^{\ell+M}}{\ell! (M-\ell)!} \,.
\label{MRichardson}
\end{align}
This transformation eliminates the first $M$ subleading terms in \eqref{sn}, producing a sequence that converges to $g_0$ significantly faster than the original one. Applying \eqref{MRichardson} to the sequences at hand, we obtain
\begin{align}
A_{150}^{\text{even}\,(10)}  = 2.000000000000000000304\ldots \, , \quad  A_{150}^{\text{odd}\,(10)}  = 2.000000000000000000121\ldots \, ,
\end{align}
which represents a substantial improvement over \eqref{AsequenceValues}. We therefore conclude, as expected, that $|A| = 2$. All the numerical results presented in the following have been obtained by applying the Richardson transform \eqref{MRichardson} to the quantity of interest. Starting from these resulting high-precision values, we can then conjecture the corresponding exact numbers based on the first few decimal digits. In particular, we apply this procedure to the following two sequences, which are instrumental for determining the remaining parameters in \eqref{anLarge}:
\begin{subequations}
\begin{align}
r_n^{\text{even}} & = \frac{a_{n+1}}{a_n}\frac{2}{n} \, ,  \quad \, n=0,2,4,6,\ldots
\label{r_even}
\\
r_n^{\text{odd}} & = \frac{a_{n+1}}{a_n}\frac{2}{n} \, ,  \quad \, n=1,3,5,7,\ldots\, \ .
\label{r_odd}
\end{align}
\label{r_sequence}
\end{subequations}
We find that the leading large-$n$ behavior of these sequences is given by
\begin{subequations}
\begin{align}
r^{\text{even}} & = 1+\frac{3}{n}-\frac{1}{2n^2}-\frac{21}{n^3}-\frac{393}{8 n^4}-\left(72\zeta_3+\frac{1113}{8}\right)\frac{1}{n^5}-\left(444 \zeta_3+\frac{11163}{16}\right)\frac{1}{n^6} \nonumber \\
& -\left(2304 \zeta_3+720 \zeta_5+\frac{40509}{8}\right)\frac{1}{n^7} -\left(14997 \zeta_3+10890 \zeta_5+\frac{5305437}{128}\right)\frac{1}{n^8} + O\left(\frac{1}{n^{9}}\right)\, \ , \\
r^{\text{odd}} & = 1-\frac{1}{n}-\frac{1}{2n^2}+\frac{1}{n^3}-\frac{129}{8
n^4}-\left(24 \zeta_3+\frac{181}{8}\right)\frac{1}{n^5}-\left(108\zeta_3+\frac{4243}{16}\right)\frac{1}{n^6}\nonumber\\
&-\left(804 \zeta_3+360 \zeta_5+\frac{15503}{8}\right)\frac{1}{n^7}-\left(6129 \zeta_3+4770 \zeta_5+\frac{2211405}{128}\right)\frac{1}{n^8}+O\left(\frac{1}{n^{9}}\right)\, \ .
\end{align}
\label{r_sequence_largen}
\end{subequations}
Then, after performing algebraic manipulations analogous to those leading to \eqref{eq:Amodule}, we find that the only choice consistent with the asymptotic expansions \eqref{r_sequence_largen} is to set $b_1 = b_2 = 0$ in \eqref{anLarge}. This choice corresponds to the presence of two logarithmic branch cuts.
Moreover, one can also show that the expansions \eqref{r_sequence_largen} force us to retain only the simple pole located along the positive real axis.

Then, by exploiting once again the \eqref{r_sequence_largen} and performing some algebra we determine as well the coefficients\footnote{We observe that, although the coefficients $c_i^{(2)}$ will not appear in the expressions for the non-perturbative corrections to $\widetilde{\mathcal{I}}^{(0)}$, their determination is nevertheless required in order to fix the remaining set of coefficients of interest, namely $c_i^{(1)}$.
} $c_i^{(j)}$ with $i = 0, 1, 2, \ldots$ and $j = 1, 2$  and the Stokes constant $\mathcal{S}_2$ and $\mathcal{S}_1$. Specifically, we find that
\begin{align}
\mathcal{S}_2 = - \frac{\mathcal{S}_1}{2} = -4i    
\end{align}
and for the first few values of $i$, we obtain
\begin{subequations}
\begin{align}
c_i^{(2)} & = \left\{1,-1,\frac{3}{32},-\frac{3}{16},-\frac{969}{2048},-\frac{2439}{2048},-\frac{228933}{65536},-\frac{39330
   9}{32768},\cdots\right\}\, \ ,
\label{c2_sequence}   \\[0.5em]
c_{i}^{(1)} & = \left\{\frac{5}{4},\frac{61}{32},\frac{213}{128},\frac{3 \zeta_3}{4}+\frac{2715}{2048},\frac{9 \zeta_3}{16}+\frac{13671}{8192},\frac{153 \zeta_3}{128}+\frac{45
   \zeta_5}{32}+\frac{173961}{65536},\cdots\right\}\, \ , 
\label{c1_sequence}   
\end{align}
\end{subequations}
where the dots stand for coefficients corresponding to higher values of the index $i$. 
Interestingly, we observe that only the coefficients $c_i^{(1)}$ depend on the odd Riemann zeta values. Then, using the result \eqref{c1_sequence} above, we finally find the first non-perturbative contribution to the asymptotic series \eqref{I0strong}, which is given by 
\begin{align}
\frac{\text{e}^{-2\sqrt{\lambda}}}{\sqrt{\lambda}}\left[8+(8i)\sum_{n \geq 0}c_n^{(1)}\left(\frac{1}{\sqrt{\lambda}}\right)^{n}\right]\, \ .
\label{I0N4}
\end{align}
\subsection{The next to planar term}
\label{subsec:NextToPlanarTerm}
The large-$\lambda$ expansion of the next-to-planar contribution \eqref{I1tilde} contains both terms with an infinite asymptotic expansion, as in the planar case, and contributions with a truncated strong coupling expansions, analogous to those discussed in Section~\ref{sec:N2theory}. The latter originates from the constant term $+1$ inside the square brackets in the third line of \eqref{K1tilde}, which gives a term proportional to \eqref{K0}, as well as from the first line of \eqref{K1tilde}. This term, however, has not been encountered before. As shown in Appendix \ref{app:K0hat}, it can nevertheless be rewritten in terms of the functions \eqref{integralK0I} and \eqref{integralK1I}, allowing its strong-coupling expansion to be straightforwardly obtained by following the procedure outlined in Section~\ref{sec:N2theory}.

For this reason, we henceforth focus only on the terms with an infinite asymptotic expansion, to which we apply the same procedure used for the planar contribution. An inspection of \eqref{I1tilde} shows that this still requires the use of the expansions \eqref{Q10} and \eqref{Q11}, together with the asymptotic expansion of
\begin{align}
\mathcal{Q}_{(0,0)}(\lambda)
= \frac{\sqrt{\lambda}}{2\pi}\int_0^{\infty}\!dt\,
\frac{\mathrm{e}^t\, t^3}{(1-\mathrm{e}^t)^2}
\frac{1}{4\pi^2+t^2}\,
J_{0}\!\left(\frac{t\sqrt{\lambda}}{2\pi}\right)
J_{0}\!\left(\frac{t\sqrt{\lambda}}{2\pi}\right)\, .
\label{Q00}
\end{align}
This expression, which involves two $J_0$ Bessel functions and originates from the third line of \eqref{K1tilde}, admits a straightforward strong-coupling expansion. One finds
\begin{align}
\mathcal{Q}_{(0,0)}(\lambda)\,
\underset{\lambda \rightarrow \infty}{\sim}\, \frac{\lambda}{4\pi^{3/2}}
\sum_{s=1}^{\infty}
\frac{(2\pi)^{2s}}{\lambda^{s}}
\frac{1}{\Gamma\!\left(\tfrac{3}{2}-s\right)^{3}\Gamma(s)}
\,\mathcal{J}(s)\, ,
\end{align}
where
\begin{align}
\mathcal{J}(s)
= (-1)^{s-1} 4^{1-s} \pi^{-2(s-1)}
\left(
\sum_{i=1}^{s-2} i\,\zeta_{2i+1}
-\frac{1}{2}\,\delta_{s,1}
+\frac{\pi^{2}}{12}
-\frac{1}{4}
\right) .
\end{align}
Putting together the different contributions, we obtain the asymptotic expansion of the next-to-planar term,
\begin{align}
\widetilde{\mathcal{I}}^{(1)}(\lambda)
\underset{\lambda\to\infty}{\sim}
\sqrt{\lambda}\sum_{n=0}^{\infty} b_n\,\lambda^{-n/2} \, ,
\label{I1tildestrong}
\end{align}
with, for instance, the first coefficients given by
\begin{align}
b_0 = \frac{1}{8}\, , \qquad
b_1 = 0\, , \qquad
b_2 = -\frac{3}{64}\, , \qquad
b_3 = -\frac{3}{32}\bigl(1+4\zeta_3\bigr)\, .
\end{align}
Higher-order coefficients can be determined straightforwardly. \\
After expanding up to a very high order we construct a diagonal Padé-Borel approximant of order $70$ for the Borel transform of the asymptotic expansion \eqref{I1tildestrong}.  
This numerical analysis strongly suggests that the points $ t=\pm 2 $ in the Borel plane are accumulation points of the poles of the Borel--Pad\'e approximant. We are therefore led to consider an ansatz analogous to \eqref{anLarge}, supplemented by the presence of two double poles located at $ t=\pm 2 $.

To constrain the parameters entering this ansatz and to determine the nature of the associated branch cuts using the numerical methods introduced in Section~\ref{subsec:PlanarTerm}, we begin by studying the large-$n$ behavior of the sequence analogous to \eqref{r_sequence} in the present case. For even values of $n$, we find

\begin{align}
\frac{2}{n}\frac{b_{n+1}}{b_n} \underset{n>>1}{\sim} & 1+ \frac{4}{n}+ \frac{7}{2}\frac{1}{n^2}-\frac{35}{2}\frac{1}{n^3}-\frac{533}{8}\frac{1}{n^4}-\left(\frac{823}{4}+72\zeta_3\right)\frac{1}{n^5}-\left(\frac{14455}{16}+516\zeta_3\right)\frac{1}{n^6} \nonumber \\
& -\left(2820 \zeta_3+720 \zeta_5+\frac{95473}{16}\right)\frac{1}{n^7} +O\left(\frac{1}{n^8}\right)\, \ ,
\label{bn_RatioEven}  
\end{align}
while for $n$ odd, we have
\begin{align}
\frac{2}{n}\frac{b_{n+1}}{b_n} \, \underset{n>>1}{\sim} \, & 1-\frac{1}{2}\frac{1}{n^2}+\frac{1}{2}\frac{1}{n^3}-\frac{125}{8}\frac{1}{n^4}-\left(\frac{153}{4}+24\zeta_3\right)\frac{1}{n^5}-\left(\frac{4855}{16}+132\zeta_3\right)\frac{1}{n^6} \non \\
& -\left(936 \zeta_3+360 \zeta_5+\frac{35861}{16}\right)\frac{1}{n^7} + O\left(\frac{1}{n^8}\right)\, \ .
\label{bn_RatioOdd}
\end{align}
Using these results, a lengthy but straightforward computation allows one to fix the parameters entering the ansatz for the large-$n$ behavior of the coefficients $b_n$, which then reduces to\footnote{We report in Appendix \ref{app:PoleContributions} the explicit computation of the double pole contribution.}

\begin{align}
b_n \, & \, \underset{n >> 1}{\sim} \frac{\mathcal{S}_1}{2\pi i \, 2^n} \Biggl[ 
\sum_{j \ge 0} d_j^{(1)} 2^j \, \Gamma(n-j) 
+ \frac{d}{2} \, \Gamma(n+1) 
- \frac{1}{4} \, \Gamma(n+2)
\Biggr] \nonumber\\
&\quad + \frac{\mathcal{S}_2}{2\pi i \, (-2)^n} \Biggl[
\sum_{j \ge 0} d_j^{(2)} (-2)^j \, \Gamma(n-j) 
- \frac{1}{2} \, \Gamma(n+1)
\Biggr]\, ,
\end{align}
%This expression encodes two logarithmic branch cuts, together with a simple and a double pole at $t=2$, and a simple pole at $t=-2$ ,while
where $\mathcal{S}_1$ and $\mathcal{S}_2$ are two Stokes constants. After some algebra we find
\begin{align}
\mathcal{S}_2 = \frac{\mathcal{S}_1}{2} = i  \, , \quad d=-\frac{1}{4}\, \ , 
\label{S2,d}
\end{align}
as well as the first coefficients $d_j^{(i)}$ for $i=1,2$, namely
\begin{subequations}
\begin{align}
d_j^{(2)} & = \left\{-\frac{1}{2},\, \frac{3}{32},\,-\frac{15}{64},-\frac{585}{2048}, \frac{423 \zeta_3}{32}+\frac{27 \zeta_5}{8}+\frac{562983}{20480}\right\}\, ,
\label{d2set}
\\
d_j^{(1)} & = \left\{-\frac{41}{32},\, -\frac{213}{128},-\frac{3 \zeta_3}{4}-\frac{4419}{2048},-\frac{81 \zeta_3}{8}-\frac{9 \zeta_5}{4}-\frac{886439}{40960}\right\}\, \ .
\label{d1set}
\end{align}
\label{dset}
\end{subequations}
Finally, putting together the different contributions we determine the first non-perturbative corrections to the asymptotic expansion of \eqref{I1tilde}, which reads
\begin{align}
& \text{e}^{-\sqrt{\lambda}}\left[\sqrt{\frac{\pi}{2}}\left(\frac{1}{4}\frac{1}{\lambda^{7/4}}-\frac{21}{32}\frac{1}{\lambda^{5/4}}+\left(\frac{243+64\pi^2}{1536}\right)\frac{1}{\lambda^{3/4}}+O\left(\frac{1}{\lambda^{1/4}}\right)\right)\right] \nonumber \\
&+
\text{e}^{-2\sqrt{\lambda}}\left[-\frac{2}{\lambda}+\frac{1}{2\sqrt{\lambda}}+(2i)\sum_{n \geq 0}d_{n}^{(1)}\left(\frac{1}{\sqrt{\lambda}}\right)^{n+1
}  \right] + O\left(\text{e}^{-3\sqrt{\lambda}}\right)
\label{I1NonPerturbative}
\end{align}
We observe that the first line of \eqref{I1NonPerturbative} arises solely from the use of the method  developed in Section \ref{sec:N2theory}, originating from the expansions of $\mathcal{K}^{(0)}$ \eqref{K0final} and $\widehat{\mathcal{{K}}}^{(0)}$ \eqref{K0hatfinal}. Contributions associated with the coefficients $d_n^{(1)}$ appear only in the second line.

\section{Conclusions}
\label{sec:Conclusions}
In this work, we provided a systematic study of the non-perturbative corrections at strong coupling 
to the integrated correlator \eqref{eq:IntegratedCorrelatorI} in two theories with gauge group $Sp(N)$ 
that differ in the amount of supersymmetry. In the $\mathcal{N}=2$ case, as reviewed in Section \ref{Sec: Toda}, 
the Toda chain equation allows one to express the large-$N$ coefficients algorithmically in terms of only three functions: 
$\partial_{\lambda}F^{(1)}$, $\partial_{\lambda}\mathcal{M}^{(0)}$, and $\mathcal{K}^{(0)}$. These coefficients can then be 
resummed in terms of the modified Bessel functions $K_\nu$, as shown in Section \ref{sec:N2theory}, 
which provides direct access to the exponentially small non-perturbative corrections and yields the expansions 
\eqref{StrongI0} and \eqref{I1I2}. The full exact expressions are reported in the ancillary \texttt{Mathematica} file.  Both the resulting expansions and the methodology employed constitute some 
of the main results of this work. For the $\mathcal{N}=4$ case, we were not able to identify a closed-form resummation of the perturbative strong-coupling expansions nor was it possible to rely on a similarly direct algorithmic recursion like the Toda chain equation. Consequently, the determination of the non-perturbative sector required a resurgent numerical analysis \cite{marino_resurgence_course,Aniceto:2018bis}: by generating high-order perturbative coefficients, we investigated the singularity structure of the Borel plane, to extract the leading exponentially small corrections for both the planar $\widetilde{\mathcal{I}}^{(0)}$ and next-to-planar $\widetilde{\mathcal{I}}^{(1)}$ contributions, as detailed in equations \eqref{I0N4} and ~\eqref{I1NonPerturbative}.

Since both theories here analyzed admit a gravity dual, it is natural to investigate the holographic counterpart of the integrated correlators here considered. In both cases, the holographic dual is of the form $\text{AdS}_5 \times S^5/\Gamma$, where $\Gamma$ is a discrete group acting on the internal space \cite{Ennes:2000fu}. Before considering the holographic interpretation of the non-perturbative corrections derived in this work we find it pedagogical to review the case of the v.e.v. of a Wilson loop. It turns that, in the 't Hooft limit and at tree level, this quantity in holography is computed by minimizing the classical action of a fundamental string ending on the loop at the AdS boundary \cite{Maldacena:1998im}. Moreover, also its non-perturbative corrections admit an holographic interpretation since arise by the expansion around the other saddle point of the string action \cite{Drukker:2006ga}. \\
In \cite{Pufu:2023vwo} it was shown that for the $\mathcal{N}=4$ SYM, with $SU(N)$ gauge group, the integrated correlator involving a single Wilson line is dual to the scattering amplitude of a massless string mode off an extended fundamental string, which is dual to the Wilson line and stretches across $\text{AdS}_5$, together with two massless closed string modes originating from insertions of the moment map operators at the $\text{AdS}_5$ boundary. To support this identification the authors of \cite{Pufu:2023vwo} employ the effective Green-
Schwarz action \cite{GREEN1984367,GREEN1984475,Kallosh:1998ji,Drukker:2000ep} for the fundamental string 
\begin{align}
S_{\text{F}_1} = -T_{\text{F}_1} \int d^2\sigma \sqrt{-\text{det}\, G_{\mu\nu}(X)\partial_{a}X^{\mu}\partial_{b}X^{\nu}} \, + \, \cdots, \, \
\label{SF1action}
\end{align}
where $T_{\text{F}_1}$ is the string tension, $G_{\mu\nu}$ the target space metric, $X^{\mu}$ ($\mu=0,\cdots,9$) the target-space coordinates, and $\sigma^{a}$ ($a=0,1$)  the worldsheet coordinates. Finally, the ellipses encodes the coupling to the NS-NS 2-form $B_{\mu\nu}$ and higher derivative corrections. Expanding the action \eqref{SF1action} in static gauge, $X^{0}=\sigma^{0}$ and  $X^{1}=\sigma^{1}$, around a background metric $G^{(0)}_{\mu\nu}$, with $G_{\mu\nu} = G^{(0)}_{\mu\nu} + h_{\mu\nu}$ and $h_{\mu\nu}$ representing a bulk graviton, yields both the graviton–worldsheet coupling and interaction vertices involving the scalars $X^{i}$ (with $i=2,\cdots,9$) living on the F1 worldsheet. We expect  that the identification via the AdS/CFT correspondence between line-defect integrated correlators and the scattering amplitudes mentioned above can be naturally extended also to the $Sp(N)$ theories considered here. The difference due to orbifold/orientifold projections should be encoded in a different background geometry $\widetilde{G}^{(0)}_{\mu\nu}$, while the worldsheet fluctuations must satisfy specific parity conditions under the further symmetries imposed by the discrete group $\Gamma$, effectively modifying the spectrum of scattering modes compared to the $SU(N)$ case. 
In this set up we are lead to conjecture that the non-perturbative corrections derived in this work could by obtained by the expansion around different saddle points of the analogous of the string action \eqref{SF1action} for the $Sp(N)$ SCFTs.

Finally, a key result of this work is the introduction of a new strong-coupling resummation method, which is particularly effective when the perturbative expansion truncates and  complementary to Cheshire Cat resurgence. Indeed rather than reconstructing non-perturbative effects from asymptotic data, this novel method yields closed analytic expressions that make exponentially suppressed contributions at strong coupling manifest. An explicit comparison with the Cheshire cat resurgence
has been performed in Appendix \ref{App:Cheschire}. Specifically, in this work, we applied this method to observables that are represented by integrals involving a single Bessel $J$ function that also appear in the computation of integrated correlators in others $\mathcal{N}=2$ SCFTs, as for example the $\mathbf{D}$-theory \cite{Billo:2024ftq, DeLillo:2025stg}. There a crucial role is played by the coefficients $\textsf{Z}_k^{(p)}$, defined as 
\begin{align}
    \textsf{Z}_k^{(p)}= \int_0^{\infty}dt\; \frac{\text{e}^t \;t^p}{(\text{e}^t-1)^2} \; J_k \left(\frac{t\sqrt{\lambda}}{2\pi} \right)\;.
\end{align}
that are the generalization of \eqref{M0old}. Therefore the method developed in Section \ref{sec:N2theory} can be used to extract non-perturbative corrections in those cases as well. Moreover, an important open problem is to extend this resummation strategy to more general integrals, such as those involving two Bessel $J$ functions appearing in the matrix model evaluation of 2-point and 3-point functions among scalar chiral operators \cite{Beccaria:2020hgy,Billo:2022xas} and correlators involving  a Wilson loop and a chiral operator \cite{Pini:2023svd}. 
Works along these lines are in progress and would significantly enlarge the range of observables amenable to exact strong-coupling resummation.

\vskip 1cm
\noindent {\large {\bf Acknowledgments}}
\vskip 0.2cm
We are very grateful to Marialuisa Frau for carefully reading the manuscript and to Zoltan Bajnok, Marco Billò, Gerald V. Dunne, Marialuisa Frau, Francesco Galvagno, Alberto Lerda and Paolo Vallarino  for many relevant discussions.

This research is partially supported by
the INFN project ST\&FI
``String Theory \& Fundamental Interactions''.
The work of AP is supported  by the Deutsche Forschungsgemeinschaft (DFG, German Research Foundation) via the Research Grant ``AdS/CFT beyond the classical supergravity paradigm: Strongly coupled gauge theories and black holes” (project number 511311749). 

\vskip 1cm

\appendix

\section{Comparison with Cheshire cat resurgence}
\label{App:Cheschire}
In this Appendix we want to retrieve the non-perturbative strong coupling expansion of $\partial_{\lambda}^2\left(\lambda F^{(1)} \right)$, reported in equation ($6.16$) of \cite{Beccaria:2021ism} \footnote{Note that in \cite{Beccaria:2021ism} a SU($N$) model is considered and its planar component of the free energy is one half of that of the Sp($N$) one analyzed here.}, using the Cheshire cat resurgence method, following \cite{Brown:2025huy}.
Starting from \eqref{F1} we get
\begin{align}
    \frac{d^2}{d\lambda^2}\big( \lambda F_1 \big)= \frac{\log 2}{ \pi^2} -\frac{2}{\pi \sqrt{\lambda}}\int_0^{\infty}dt\; \frac{\text{e}^t}{(\text{e}^t +1)^2}\;J_1\left(\frac{t \sqrt{\lambda}}{\pi}\right) \;.
    \label{d2F1}
\end{align}
Exploiting the Mellin-Barnes representation of the Bessel $J$ function and performing the integral over $t$ using the integral representation of the Dirichlet $\eta$ function, \eqref{d2F1} can be rewritten as
\begin{align}
    \frac{d^2}{d\lambda^2}\big( \lambda F_1 \big)= \frac{\log 2}{ \pi^2} -\frac{2}{\pi \sqrt{\lambda}} \int_{-i \infty}^{i \infty} \frac{ds}{2\pi i} \frac{\Gamma(-s)\Gamma(2s+2)}{\Gamma(s+2)} \eta(2s+1)\left(\frac{\sqrt{\lambda}}{2\pi} \right)^{2s+1}  \;.
    \label{IMB}
\end{align}
Then, the perturbative strong coupling expansion can be obtained by closing the integration contour in the anti-clockwise way, getting
\begin{align}
    \frac{d^2}{d\lambda^2}\big( \lambda F_1 \big) \underset{\lambda \to \infty}{\sim} \frac{\log 2}{\pi^2} -\frac{1}{2\lambda}\;. 
    \label{SCI}
\end{align}
Since this asymptotic expansion has a finite number of terms it is not possible to employ the standard resurgence techniques to extract the non-perturbative corrections. Therefore we deform the quantity under consideration in order to have a perturbative strong coupling expansion with infinite terms introducing the auxiliary function
\begin{align}
   \mathcal{T}(b) =\frac{\log 2}{\pi^2} -\frac{2}{\pi \sqrt{\lambda}} \int_{-i \infty}^{i \infty} \frac{ds}{2\pi i} \frac{\Gamma(-s)\Gamma(2s+2)}{\Gamma(s+b)} \eta(2s+1)\left(\frac{\sqrt{\lambda}}{2\pi} \right)^{2s+1} \;,
    \label{SC1B}
\end{align}
where  $b \in \mathfrak{C}$ is a deformation parameter. For $b=2$ \eqref{SC1B} clearly reduces to \eqref{IMB}. Closing the integration contour in the anti-clockwise way and picking up the poles located at $s=-n$, with $n=1,2,...$, we are able to get the full perturbative strong-coupling expansion that, after some algebra, reads
\begin{align}
   \mathcal{T}(b) \underset{\lambda \to \infty}{\sim} \mathcal{T}^{(\text{p})}(b) = \frac{\log 2}{ \pi^2}  -\frac{1}{2 \lambda  \Gamma (b-1)} + \frac{\sin(\pi b)}{\pi^3}\sum_{n=2}^{\infty}  \left(4^n-1\right) (2 n-1)  \zeta (2 n) \Gamma (n)  \Gamma (-b+n+1) \lambda ^{-n}\;.
   \label{SCIb}
\end{align}
It is worth noting that if we set $b=2$ we correctly recover \eqref{SCI}. Now we define the modified Borel transform \cite{Arutyunov:2016etw}
\begin{align}
    \widetilde{\mathbf{B}}: \quad \quad \sum_{n=1}^{\infty}c_n \, \lambda^{-n} \longrightarrow \phi(w,b)= \sum_{n=1}^{\infty}\frac{2c_n}{\zeta(2n)\;\Gamma(2n+1)}\;(2w)^{2n} \;,
    \label{Btilde}
\end{align}
that differs from the standard Borel transform for the division by $\zeta(2n)$. In the case of the coefficients $c_n$ reported in \eqref{SCIb} it reads
\begin{align}
     \phi(w,b)= \phi_1(w,b)+ \phi_2(w,b) + \phi_3(w,b)\;,   \label{phidec}
\end{align}
where 
\begin{subequations}
\begin{align}
  &  \phi_1(w,b)=  -\frac{4 \sin(\pi b)}{\pi^3}\Gamma (2-b) \bigg(2 w^2  \, _2F_1\left(1,2-b;\frac{3}{2};w^2\right)-w^2  \, _3F_2\left(1,1,2-b;\frac{3}{2},2;w^2\right)\bigg) \\
  & \phi_2(w,b)=  -\frac{4 \sin(\pi b)}{\pi^3}\Gamma (2-b) \bigg(-8 w^2  \, _2F_1\left(1,2-b;\frac{3}{2};4 w^2\right)+4 w^2 \, _3F_2\left(1,1,2-b;\frac{3}{2},2;4 w^2\right)\bigg) \\
  & \phi_3(w,b)= -\frac{12 w^2 \Gamma (2-b)}{\pi ^3}\;.
\end{align}
\end{subequations}
Then we introduce the directional Borel summation of \eqref{SC1B} as 
\begin{align}
   \mathcal{S}_{\theta}[\mathcal{T}^{(\text{p})}(b)]= \frac{\log 2}{ \pi^2} -\frac{1}{2 \lambda  \Gamma (b-1)} + \sqrt{\lambda} \int_0^{e^{i\theta}\infty} \frac{dw}{4 \sinh^2(\sqrt{\lambda}w) } \;\phi(w,b) \;,
    \label{transform}
\end{align}
which defines an analytic function for $\lambda >0$ when $\theta \in (-\pi/2,0) \; \cup \; (0, \pi/2) $. In particular, in \eqref{transform} it is not possible to integrate along the real line $\theta=0$ since $\phi_1$ has a branch cut for $w \in [1, + \infty)$ while $\phi_2$ for $ w \in [1/2, + \infty)$ and this fact implies that \eqref{SC1B} is not Borel summable. In order to dodge the singularity located in $\theta=0$ we introduce the two lateral summations 
\begin{align}
    \mathcal{S}_{\pm}[\mathcal{T}^{(p)}(b)]= \lim_{\theta \to 0^{\pm}} \mathcal{S}_{\theta}[\mathcal{T}(b) ] \;,
\end{align}
that in principle provide two different analytical continuations of \eqref{SC1B} and their difference precisely encodes the exponentially small non-perturbative terms. Then, to obtain a real and unambiguous resummation of \eqref{SC1B} we take an average of the two lateral resummations, referred to as  ``median resummation''  \cite{article}, namely
\begin{align}
    &\mathcal{S}_{\text{med}}[\mathcal{T}^{(p)}(b)]= \mathcal{S}_{\pm}[\mathcal{T}^{(p)}(b)] \mp \frac{1}{2} \Delta[\mathcal{T}^{(p)}(b)]\;,
    \label{Smed}
\end{align}
with 
\begin{align}
    \Delta[\mathcal{T}^{(p)}(b)] \equiv   ( \mathcal{S}_+-\mathcal{S}_-)[\mathcal{I}(b)]=\sqrt{\lambda} \int_0^{\infty} \frac{dw}{4 \sinh^2(\sqrt{\lambda}w) } \;\text{Disc}\;\phi(w,b)\;,
    \label{Delta}
\end{align}
which encodes the exponentially small non-perturbative corrections.
Exploiting
\begin{align}
   & \mathrm{Disc}\, {}_2F_{1}(a,b;c;z) \;= \nonumber \\ 
   \;&
\,
\frac{2\pi i \;\Gamma(c)}{\Gamma(a)\,\Gamma(b)\,\Gamma(c-a-b+1)}\,
z^{1-c}\,(z-1)^{\,c-a-b}\; \, _2F_1(1-a,1-b;-a-b+c+1;1-z) \;,
\label{2F1disc}
\end{align}
and \cite{Brown:2025huy}
\begin{align}
   & \mathrm{Disc}\, {}_3F_{2}\left(1,1,2-b;\frac{3}{2},2;z\right) \;= \nonumber \\ 
   \;&
 -\,i\,\sin(\pi b)\,
 \sqrt{\pi}\, z^{-\tfrac{1}{2}}\,
 (z - 1)^{\,b - \tfrac{1}{2}}\,
 \frac{\Gamma(b - 1)}{\Gamma\left(b+\frac{1}{2} \right)}\,
 _2F_1\!\left(1,\, b;\, b + \tfrac{1}{2};\, 1 - z\right)
 \;,
\end{align}
we express the discontinuity accross the real line of \eqref{phidec} in \eqref{Delta} as 
\begin{align}
    \text{Disc}\;\phi(w,b)= -2i\;\sin(\pi b) \;\bigg(\Delta_1(w,b) + \Delta_2(w,b)\bigg) \;,
    \label{Disc}
\end{align}
with 
\begin{subequations}
\begin{align}
    \Delta_1(w,b) =  &  \frac{2w}{\pi^{3/2}} \bigg[-\frac{\left(w^2-1\right)^{b-\frac{1}{2}}}{\Gamma\left(b+\frac{1}{2} \right)} \, _2 F_1\left(1,b;b+\frac{1}{2};1-w^2\right)+\frac{2 \left(w^2-1\right)^{b-\frac{3}{2}}}{\Gamma \left(b-\frac{1}{2}\right)}\bigg] \;\\
    \Delta_2(w,b)= & \frac{2w}{\pi^{3/2}} \bigg[2 \frac{\left(4 w^2-1\right)^{b-\frac{1}{2}}}{\Gamma \left(b +\frac{1}{2} \right)} \, _2F_1\left(1,b;b+\frac{1}{2};1-4 w^2\right)-\frac{4 \left(4 w^2-1\right)^{b-\frac{3}{2}}}{\Gamma \left(b-\frac{1}{2}\right)}\bigg] \;.
\end{align}
\end{subequations}
Using \eqref{Disc} we rewrite \eqref{Delta} as 
\begin{align}
&\mathcal{S}_{\text{med}}[\mathcal{T}^{(p)}(b)]=\mathcal{S}_{\pm}[\mathcal{T}^{(p)}(b)] \;\pm\; i \sin(\pi b)\; \big(\mathcal{T}^{\text{(np)}}_1(b, \lambda) +\mathcal{T}^{\text{(np)}}_2(b, \lambda) \big)
\label{Smed2}
\end{align}
with 
\begin{subequations}
\begin{align}
 &\mathcal{T}^{\text{(np)}}_1(b, \lambda) =  \sqrt{\lambda} \int_{1}^{\infty} \frac{dw}{4 \sinh^2(\sqrt{\lambda}w) } \;\; \Delta_1(w,b) \label{T1}   \\
 &\mathcal{T}^{\text{(np)}}_2(b,\lambda) =  \sqrt{\lambda} \int_{\frac{1}{2}}^{\infty} \frac{dw}{4 \sinh^2(\sqrt{\lambda}w) } \;\; \Delta_2(w,b) \label{T2}\;.
\end{align}
\end{subequations}
In \eqref{Smed2}, following \cite{Brown:2025huy}, we identify the factor $\sin(\pi b)$ as the imaginary part of the transseries parameter $\sigma(b)$ that we suppose to be 
\begin{align}
    \sigma(b)= \text{e}^{i \pi b}\;.
\end{align}
Then, we claim that the complete transseries representation of \eqref{SCI} is given by 
\begin{align}
    \mathcal{T}(b) = \mathcal{S}_{\pm}[\mathcal{T}^{(p)}(b)]+ \sigma(b) \big(\mathcal{T}^{\text{(np)}}_1(b, \lambda) +\mathcal{T}^{\text{(np)}}_2(b, \lambda) \big)\;.
\end{align}
Now we specialize the above transseries to the physical value $b=2$ in order to extract the non-perturbative corrections 
\begin{align}
    \mathcal{T}^{(np)}\equiv \mathcal{T}_1^{(np)}+\mathcal{T}_2^{(np)}\;,
\end{align}
with
\begin{align}
    \mathcal{T}_i^{(np)} \equiv   \mathcal{T}_i^{(np)} (2, \lambda) \quad \quad \quad i=1,2\;.
\end{align}
Let's analyze in detail the computation of the integral \eqref{T1} at $b=2$. After the use of the identity \eqref{expid} to rewrite the integration measure and the shift of the integration contour $w \to w+1$, we get 
\begin{align}
    \mathcal{T}^{(np)}_1=  \frac{4\sqrt{\lambda}}{\pi^2}\sum_{n=1}^{\infty}n \; \text{e}^{-2n\sqrt{\lambda}}\int_{0}^{\infty}dw  \;\text{e}^{-2nw\sqrt{\lambda}} \;\left( (w+1)\sqrt{w (w+2)} + \sinh^{-1}\left(\sqrt{w (w+2)}\right) \right)\;.
\end{align}
Then we Taylor expand the integrand around $w=0$, we compute the resulting integral by means of the identity 
\begin{align}
       \int_0^{\infty }  \text{e}^{-2 \sqrt{\lambda } n w} w^z\, dw = \frac{\Gamma(z+1)}{(2n \sqrt{\lambda})^{z+1}}\;,
\end{align}
and, after the rescaling $n \to 2n$, we get 
\begin{align}
    \mathcal{T}_1^{(np)} = \frac{2}{\pi^2} \sqrt{\lambda}\sum_{n=1}^{\infty}n\;\text{e}^{-2n\sqrt{\lambda}}\sum_{k=0}^{\infty} \frac{(-1)^k  \left(4 k^2+3\right)  \Gamma \left(k-\frac{3}{2}\right)}{2^{k+\frac{1}{2}}\;\pi^{5/2} \; (2k+1)\Gamma (k+1)} \frac{\Gamma(k+\frac{3}{2})}{(n \sqrt{\lambda})^{k+ \frac{3}{2}}} 
    \label{T1np}\; \;.  
\end{align}
Following similar steps for $\mathcal{T}_2$ in \eqref{T2} at $b=2$, we get 
\begin{align}
\mathcal{T}_2^{(np)} = -\frac{2}{\pi^2}\sqrt{\lambda}\sum_{n=1}^{\infty} n \;\text{e}^{-n\sqrt{\lambda}} \sum_{k=0}^{\infty}  \frac{ (-1)^k \left(4 k^2+3\right)  \Gamma \left(k-\frac{3}{2}\right)}{2^{k+\frac{1}{2}}\pi ^{5/2} (2 k+1) \Gamma (k+1)}\; \frac{\Gamma(k+\frac{3}{2})}{(n \sqrt{\lambda})^{k+ \frac{3}{2}}}
\label{T2np}
\end{align}
Finally, summing \eqref{T1np} and \eqref{T2np} we end up with the following non-perturbative expansion
\begin{align}
    \mathcal{T}^{(np)} = -\frac{2\sqrt{2}}{\pi^2}\sum_{n=0}^{\infty}\text{e}^{-(2n+1)\sqrt{\lambda}}\sum_{k=0}^{\infty}\frac{(-1)^k(k^2+\frac{3}{4})\Gamma(k-\frac{3}{2})}{2^k\sqrt{\pi}\Gamma(k+1)}\frac{\Gamma(k+\frac{1}{2})}{((2n+1)\sqrt{\lambda})^{k+1/2}} \;.
\end{align}
This result perfectly matches that obtained in equation ($6.16$) of \cite{Beccaria:2021ism}, once taking into account the different normalization of the free energy. It is worth noting that the Cheshire Cat resurgence only captures the exponentially small corrections to the strong coupling expansions considered, whereas the method described in Section \ref{sec:N2theory} provides an exact-in- $\lambda$ unintegrated expression for the various observables.

\section{Closed-form expression for $\widehat{\mathcal{K}}^{(0)}$}
\label{app:K0hat}

In this Appendix, using the method outlined in Section~\ref{sec:N2theory}, we derive a closed-form expression for the function
\begin{align}
\widehat{\mathcal{K}}^{(0)} =  4\pi\int_0^{\infty}dt\,\frac{\mathrm{e}^{t}\,t^2}{(1-\mathrm{e}^{t})^2}\frac{1}{\pi^2+t^2}\,\mathcal{B}(2t)\,,
\label{K0hat}
\end{align}
where $\mathcal{B}(t)$ was defined in~\eqref{B}. This expression is required for the evaluation of the next-to-planar contribution to the $\mathcal{N}=4$ SYM integrated correlator performed in Section~\ref{subsec:NextToPlanarTerm}.

We begin by observing that, after a change of variables, the expression~\eqref{K0hat} can be rewritten as
\begin{align}
\widehat{\mathcal{K}}^{(0)} = 2\pi \int_0^{\infty} dt\,\frac{\mathrm{e}^{t/2}\,t^2}{(1-\mathrm{e}^{t/2})^2} \frac{1}{4\pi^2 + t^2}\, \mathcal{B}(t)\,.
\end{align}
This motivates us to consider a deformed version, namely with a modified integration kernel, of the vacuum expectation value~\eqref{M0}. We denote this quantity by $\widehat{\mathcal{M}}^{(0)}$. In the following, it is sufficient to consider its derivative with respect to $\lambda$, which reads
\begin{align}
\partial_{\lambda}\widehat{\mathcal{M}}^{(0)} =
\frac{2}{\lambda}\int_0^{\infty} dt\,\frac{\mathrm{e}^{t/2}\,t}{(1-\mathrm{e}^{t/2})^2}
J_2 \!\left(\frac{t \sqrt{\lambda}}{2\pi}\right).
\label{derM0hat}
\end{align}

After some algebraic manipulations, one can show that the $\lambda$-derivative of~\eqref{K0hat} satisfies the differential equation~\eqref{DK0}, with the only modification that $\mathcal{M}^{(0)}$ is replaced by $\widehat{\mathcal{M}}^{(0)}$. We therefore proceed to compute the non-perturbative contributions to~\eqref{derM0hat}. Rewriting the latter as
\begin{align}
\partial_{\lambda} \widehat{\mathcal{M}}^{(0)}
&= \frac{2}{\lambda} \int_0^{\infty} dt \sum_{n=1}^{\infty}
n\, \mathrm{e}^{-nt/2}\, t \, J_2(a t) \nonumber \\
&= \frac{2}{\lambda}\sum_{n=1}^{\infty}
\left[
\frac{2n}{a^2}
-\frac{4n^2}{(n^2 + 4a^2)^{3/2}}
-\frac{2}{a^2} \frac{n^2}{\sqrt{n^2+4a^2}}
\right],
\label{derMOhat2}
\end{align}
where $a = \sqrt{\lambda}/(2\pi)$. We now analyze the three contributions appearing on the right-hand side of the expression above separately. Using the analytic continuation of the Riemann zeta function at $-1$, the first contribution gives
\begin{align}
\frac{2}{\lambda}\sum_{n=0}^{\infty}\frac{2n}{a^2}
= - \frac{4\pi^2}{3\lambda^2}\,.
\label{first}
\end{align}
For the second contribution, it is convenient to rewrite it as
\begin{align}
-\frac{8}{\lambda}\sum_{n=1}^{\infty} \frac{n^2}{(n^2+4a^2)^{3/2}}
= 4\sum_{n=1}^{\infty}
\left[-\frac{2}{\lambda} \frac{n^2}{(n^2+\tilde{a}^2)^{3/2}}\right],
\end{align}
with $\tilde{a}=2a$. The term in square brackets is identical to the first contribution on the right-hand side of~\eqref{A2} and can therefore be treated in the same way, yielding
\begin{align}
-\frac{8}{\lambda}\sum_{n=1}^{\infty} \frac{n^2}{(n^2+4a^2)^{3/2}}
&= - \frac{4}{\lambda} \int_0^{\infty}\frac{dt}{t}\, \mathrm{e}^{-4a^2 t} \nonumber \\
&\quad -\frac{16}{\lambda} \sum_{k=1}^{\infty}
\left[ K_0(4 a k \pi) - (4 a k \pi) K_1(4 a k \pi) \right].
\label{second}
\end{align}
Finally, the third contribution can be rewritten as
\begin{align}
-\frac{4}{\lambda a^2} \sum_{n=1}^{\infty} \frac{n^2}{\sqrt{n^2+4a^2}}
&= - \frac{16}{\lambda \tilde{a}^2}
\sum_{n=1}^{\infty} \frac{n^2}{\sqrt{n^2+ \tilde{a}^2}} \nonumber \\
&= 4 \left[- \frac{4}{\lambda \tilde{a}^2}
\sum_{n=1}^{\infty} \frac{n^2}{\sqrt{n^2+ \tilde{a}^2}} \right],
\label{third}
\end{align}
with $\tilde{a}=2a$. The expression in square brackets has the same form as the second contribution appearing on the right-hand side of~\eqref{A2} and can therefore be treated analogously.

Using the results~\eqref{first}, \eqref{second}, and~\eqref{third}, and performing straightforward algebraic manipulations, we arrive at
\begin{align}
\partial_{\lambda}\widehat{\mathcal{M}}^{(0)}
= \frac{4}{\lambda}
-\frac{4\pi^2}{3\lambda^2}
+ \frac{16}{\lambda} \sum_{k=1}^{\infty}
\left[
\frac{(2 \lambda k^2+1)\, K_1(2k \sqrt{\lambda})}{k \sqrt{\lambda}}
+ K_0(2 k \sqrt{\lambda})
\right] \, ,
\label{derMOhat3}
\end{align}
which is our final expression for~\eqref{derM0hat}, including the non-perturbative corrections encoded in the modified Bessel functions $K_\nu(2k\sqrt{\lambda})$.
To obtain a closed-form expression for~\eqref{K0hat}, we now employ the analogue of the differential equation~\eqref{DK0} appropriate to the present case. Substituting~\eqref{derMOhat3} into it, we find
\begin{align}
\partial_{\lambda}\widehat{\mathcal{K}}^{(0)}
= \frac{2\,I_{1}(\sqrt{\lambda})}{\sqrt{\lambda}}
\left[
1
+ 4\sqrt{\lambda}
\sum_{k=1}^{\infty}
\left(
k\,K_{1}(2k\sqrt{\lambda})
- 2k^{2}\sqrt{\lambda}\,K_{0}(2k\sqrt{\lambda})
\right)
\right].
\end{align}

Integrating this expression by following the same procedure described in Section~\ref{sec:N2theory}, we obtain
\begin{align}
\widehat{\mathcal{K}}^{(0)}(\lambda)
&= \int_{0}^{\lambda} dq\, \partial_{q}\widehat{\mathcal{K}}^{(0)}(q) \nonumber \\
&= 4\bigl(I_{0}(\sqrt{\lambda}) - 1\bigr)
   + 4\sum_{k=1}^{\infty} \bigl( \mathcal{J}_{1}(2k,\lambda) + \mathcal{J}_{0}(2k,\lambda) \bigr)
   + (4 - \pi^{2}) \nonumber \\
&= 4\,I_{0}(\sqrt{\lambda})
+ 4\sum_{k=1}^{\infty}
\bigl( \mathcal{J}_{1}(2k,\lambda) + \mathcal{J}_{0}(2k,\lambda) \bigr)
- \pi^{2},
\label{K0hatfinal}
\end{align}
which is our final expression for $\widehat{\mathcal{K}}^{(0)}$ and
where $\mathcal{J}_0(k,\lambda)$ and $\mathcal{J}_1(k,\lambda)$ are defined in
\eqref{integralK0I} and~\eqref{integralK1I}, respectively, and encode the non-perturbative corrections.
As an additional check, we have numerically verified that this expression reproduces~\eqref{K0hat} for several values of $\lambda$.

\section{The integral \eqref{K0integral} and Dominated Convergence Theorem}
\label{App:DominatedConvergence}
In this Appendix we justify the exchange of the operations of integral and limit in equation \eqref{K0integral}, that explictly reads 
\begin{align}
    \mathcal{K}^{(0)}(\lambda)= I_0(\sqrt{\lambda}) + \int_0^{\lambda}dq\;  \lim_{N \to \infty} \left(\sum_{k=1}^N \,f_k (q) \right)\;,
\end{align}
with 
\begin{align}
    f_k(q) = I_1(\sqrt{q})\Big(k K_1(k \sqrt{q}) - k^2 \sqrt{q} K_0(k \sqrt{q})\Big).
\end{align}
Since
\begin{align}
\lim_{q \to 0^+} f_k(q) = \frac{1}{2}, \qquad
\lim_{q \to \infty} f_k(q) \sim \text{e}^{(1-k)\sqrt{q}}\;,
\label{limits}
\end{align}
it is convenient to exclude the case $k=1$. Therefore, in the following, we consider the sequence of partial sums starting from $k=2$, namely
\begin{align}
S_N(q) \equiv \sum_{k=2}^{N} f_k(q).
\end{align}
Our goal is to find an integrable function $g(q)$ such that
\begin{align}
|S_N(q)| \le g(q) \quad \forall N, \quad \text{for almost all } q \in [0,\infty).
\label{constraintOnS}
\end{align}
From the first limit in \eqref{limits}, it follows that at $q=0$ the sequence of partial sums satisfy
\begin{align}
S_N(0) = \frac{N-1}{2},
\end{align}
so that condition \eqref{constraintOnS} fails at $q = 0$, since $\lim_{N \rightarrow \infty}S_N(0) = \infty$.  However, for any arbitrarily small $\epsilon > 0$, we have
\begin{align}
\lim_{N \to \infty} S_N(\epsilon) = -\frac{3}{4},
\end{align}
and at finite $N$ the absolute value of $S_N(\epsilon)$ can always be bounded by a constant. Accordingly, we can define a dominator function
\begin{align}
g(q) =
\begin{cases}
\mathcal{C}, & 0 < q \le 1,\\[1ex]
\frac{2}{\mathrm{e}^{q}-1}, & 1 < q < \infty,
\end{cases}
\label{gDominator}
\end{align}
where $\mathcal{C}$ is an arbitrarily large but finite constant chosen to ensure that the absolute value of the sequence of partial sums is bounded for all $N$ in a neighborhood to the right of zero. Therefore, the function \eqref{gDominator} provides a valid dominator function for the sequence of partial sums. Consequently, the hypothesis of the \textit{Dominated Convergence Theorem} are satisfied, and we can interchange the limit $N \rightarrow \infty$ with the integral over $q$.

\section{Single- and double-pole contributions}
\label{app:PoleContributions}
\begin{figure}[t]
    \centering
\includegraphics[width=0.7\textwidth]{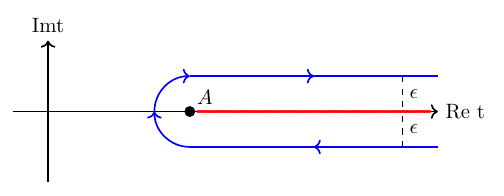}
\caption{The integration contour in the Borel plane is indicated by the blue line, the black dot represents the singularity at $
t=A$, and the red line denotes the branch cut along the positive real axis.}
\label{fig:Cammino}
\end{figure}

In this Appendix we compute the single- and double-pole contributions to the trans-series due to single and double singularities of the Borel transform. These results are employed in Section~\ref{sec:N4}. Although they are well known in the resurgence literature, we choose to present them here in order to fix our conventions regarding the choice of the integration contour in the Borel plane. Moreover, we believe that the derivation provided below may be of independent interest, as it does not rely on alien calculus~\cite{Aniceto:2018bis} and is therefore more elementary and accessible to a broader audience. Let us consider an asymptotic series of the form
\begin{align}
\varphi(z) = \sum_{n \geq 0} a_n\, z^n \,,
\label{phi}
\end{align}
and assume that its Borel transform $\widehat{\varphi}(t)$ has a simple pole singularity located on the real axis of the Borel plane at $t = A$, namely\footnote{Without loss of generality, we assume $A>0$.}
\begin{align}
\widehat{\varphi}(t) \underset{t\ \rightarrow A}{\simeq} \frac{\mathcal{S}\,a}{2\pi}\,\frac{1}{t-A}\,.
\label{SimplePole}
\end{align}
Here $\mathcal{S}$ denotes the (in general complex) Stokes constant and $a \in \mathbb{R}$. As shown, for instance  in~\cite{Dorigoni:2014hea}, the contribution of~\eqref{SimplePole} to the trans-series associated with~\eqref{phi} can be obtained by considering the integration contour depicted in Figure \ref{fig:Cammino}
. This prescription amounts to taking the difference between the Laplace integrals evaluated just above and just below the real axis, namely
\begin{align}
\left(\frac{\mathcal{S}\,a}{2\pi}\right)
\lim_{\epsilon \to 0}
\left[
\int_{0}^{\infty+i\epsilon} \frac{dt\, \mathrm{e}^{-z t}}{t-A}
-
\int_{0}^{\infty-i\epsilon} \frac{dt\, \mathrm{e}^{-z t}}{t-A}
\right].
\label{SinglePolestart}
\end{align}
A regulator $\epsilon$ has been introduced to render the expression well defined. Using the identity~\cite{DiracIdentity}
\begin{align}
\lim_{\epsilon \to 0} \frac{1}{(x \pm i\epsilon)^n}
=
\mathcal{P}\!\left(\frac{1}{x^n}\right)
\mp
\frac{i\pi (-1)^{n-1}}{(n-1)!}\,
\delta^{(n-1)}(x)\,,
\label{deltaIdentity}
\end{align}
where $\mathcal{P}$ denotes the Cauchy principal value and $\delta^{(k)}(x)$ the $k$-th derivative of the Dirac delta distribution, the expression inside the square brackets in~\eqref{SinglePolestart} can be rewritten as
\begin{align}
&\lim_{\epsilon \to 0}
\left[
\mathrm{e}^{-i\epsilon z}
\int_0^{\infty} \frac{dt\, \mathrm{e}^{-z t}}{t-A+i\epsilon}
-
\mathrm{e}^{i\epsilon z}
\int_0^{\infty} \frac{dt\, \mathrm{e}^{-z t}}{t-A-i\epsilon}
\right]
\nonumber \\
&=
\int_0^{\infty} dt\, \mathrm{e}^{-z t}
\left[
\mathcal{P}\!\left(\frac{1}{t-A}\right)
- i\pi \delta(t-A)
-
\mathcal{P}\!\left(\frac{1}{t-A}\right)
- i\pi \delta(t-A)
\right]
\nonumber \\
&=
-2\pi i \int_0^{\infty} dt\, \mathrm{e}^{-z t}\,\delta(t-A)
=
-2\pi i\, \mathrm{e}^{-A z}\,.
\end{align}
Including the overall prefactor present in~\eqref{SinglePolestart}, we conclude that the contribution of a simple pole is
\begin{align}
-\,i\,\mathcal{S}\,a\,\mathrm{e}^{-A z}\,.
\label{SinglePoleFinal}
\end{align}

In a completely analogous manner, again making use of the identity~\eqref{deltaIdentity}, one can analyze the case in which the Borel transform exhibits a double pole, namely
\begin{align}
\widehat{\varphi}(t) \underset{t \rightarrow A}{\simeq} \frac{\mathcal{S}\,a}{2\pi}\,\frac{1}{(t-A)^2}\,,
\label{DoublePole}
\end{align}
which yields the following contribution:
\begin{align}
i\,a\,z\,\mathcal{S}\,\mathrm{e}^{-A z}\,.
\label{DoublePoleFinal}
\end{align}

\printbibliography

\end{document}